\documentclass{emulateapj-rtx4}

\newcommand{\kms}{\mathrm{km\ s^{-1}}\,}

\shorttitle{The dynamics of stellar disks}
\shortauthors{Fujii et al.}

\begin{document}


\title{The dynamics of spiral arms in pure stellar disks}


\author{M. S. Fujii\altaffilmark{1}, J. Baba, T. R. Saitoh, J. Makino, and E. Kokubo}
\affil{Division of Theoretical Astronomy, 
National Astronomical Observatory of Japan,\\ 2-21-1 Osawa, Mitaka, Tokyo,
181-8588}
\email{fujii@cfca.jp}

\and

\author{K. Wada}
\affil{Graduate School of Science and Engineering, Kagoshima 
University,\\ 1-21-35, Korimoto, Kagoshima, 890-0065}
\email{wada@cfca.jp}


\altaffiltext{1}{Current address : Graduate School of Science and 
Engineering, Kagoshima University}


\begin{abstract}
It has been believed that spiral arms in pure stellar disks, especially 
the ones spontaneously formed, decay in
several galactic rotations due to the increase of stellar 
velocity dispersions. Therefore, some cooling mechanism, 
for example dissipational effects of the interstellar medium, was 
assumed to be necessary to keep the spiral arms. 
Here we show that stellar disks can maintain spiral features for several 
tens of rotations without the help of cooling, using a series of
high-resolution three-dimensional $N$-body simulations of pure stellar 
disks. We found 
that if the number of particles is sufficiently large, e.g., 
$3\times 10^6$, multi-arm spirals developed in an isolated disk can
survive for more than 10 Gyrs.  We confirmed that there is a
self-regulating mechanism that maintains the amplitude of the spiral arms.
Spiral arms increase Toomre's $Q$ of the disk, and the heating rate 
correlates with the squared amplitude of the spirals. Since the amplitude
itself is limited by $Q$, this makes the dynamical heating
less effective in the later phase of evolution.  
A simple analytical argument suggests that the
heating is caused by gravitational scattering of stars by spiral arms
and that the self-regulating mechanism in pure-stellar disks can
effectively maintain spiral arms on a cosmological timescale.
In the case of a smaller number of particles, e.g., $3\times 10^5$,
spiral arms 
grow faster in the beginning of the simulation (while $Q$ is small) and 
they cause a rapid increase of $Q$. 
As a result, the spiral arms become faint in several Gyrs. 
\end{abstract}





\section{Introduction}

The physical origin and evolution of spiral arms in disk galaxies are 
long-standing problems of galactic astronomy. The most widely known 
theory is the Lin-Shu 
hypothesis, in which spiral structures are interpreted as stationary density 
waves with a constant pattern speed in a stellar disk 
\citep{LS64,BL96}.  However, as pointed out by \citet{Toomre69}, Lin-Shu's
mechanism has a serious problem, such that the energy and 
angular momentum of the tightly wound spiral waves radially propagate
with the group velocity, and the waves are absorbed at the inner Lindblad 
resonance. Therefore, a continuous generating mechanism is necessary to 
maintain the stationary density waves, e.g., WASER mechanism; an 
outward-traveling wave is reflected and transmitted into other traveling waves
at the co-rotation resonance \citep{Mark76}.
Alternatively, it has been proposed that spiral arms grow from 
small-scale disturbances through the swing amplification mechanism 
\citep{GL65,JT66,Toomre81}. In this picture, spiral arms 
are recurrent and transient rather than stationary \citep{Toomre90,TK91}.
Previous $N$-body simulations supported the recurrent and transient spirals, 
especially for multi-arm spirals 
\citep{SC84,Sellwood00,SB02,Fuchs05,Sellwood10} and also for barred-spirals
\citep{Baba09}. 

While $N$-body simulations both with and without gas showed the recurrent 
and transient spiral arms
\citep{SC84,CF85,ET93,Bottema03,Baba09}, \citet{SC84} pointed out that 
spiral arms in pure stellar disks (i.e., without gas) disappeared in several 
galactic rotations.
They performed two-dimensional $N$-body simulations and showed that 
stars scattered by spiral arms heated up the disk (increase Toomre's $Q$ 
value), and thereby spiral arms disappeared. They argued that some dynamical 
cooling mechanism was necessary to maintain the spiral arms. They showed that 
when new stars with circular orbits (i.e., with zero velocity dispersion) 
were added 
to the stellar disk with a constant rate, the spirals were maintained for 
about 10 galactic rotations. This demonstration was based on the idea that 
stars are formed from the interstellar medium (ISM) with a small velocity 
dispersion. After their work, the effects of gas and star formation 
were investigated for cooling and also heating \citep{ET93,Bottema03}.
In addition, \citet{Bottema03} proposed that the filaments 
of gas trigger the swing amplification \citep{Toomre81} and enhance
stellar spiral arms.

However, the dynamical effect of the ISM to stellar spirals 
is not well understood yet. Recently, \citet{Baba09} showed that the 
multi-phase ISM 
in a stellar disk, in which spiral arms are self-excited, has a complicated 
velocity field, and the cold, dense gas like the giant molecular clouds 
(GMCs) has large non-circular motions. Therefore, the newly born stars are 
not necessarily dynamically cold.
Furthermore, since the mass fraction of gas in galactic disks is 
typically only $\sim$10\%, it is natural to assume that the stellar component 
controls the dynamics of the disk. Indeed,
\citet{ET93} performed two-dimensional $N$-body simulations including 
gas particles and concluded that the stellar $Q$-value controls the 
formation of spiral structures. 

Although \citet{SC84} reported that the stellar disks are heated up 
significantly in several galactic rotations, we should be careful on 
the effect of numerical artifacts.
In particular, the number of particles used in their
$N$-body simulations is only 2$\times 10^4$.
In such simulations, two-body relaxation might have significantly 
enhanced the decay of spiral arms. \citet{DT94} also performed similar
simulations for stellar disks with $m=2$ spiral arms. The number of 
particles was $5\times 10^4$. They argued that their spiral arms were 
long-lived. However the lifetime was only several galactic rotations.
Their simulations have the same problems as those of \citet{SC84}.
Therefore, three-dimensional $N$-body simulations with a large 
number of particles are necessary to investigate 
the long-term evolution of stellar disks. 
Such simulations are now feasible, thanks to the progress of
computers and numerical methods.
However, recent simulations of galactic disks have 
focused on evolution of spiral galaxies including gas 
\citep{Bottema03,Baba09}. 
There are also pure $N$-body simulations of disks with a large number 
of particles such as $5\times 10^8$ \citep{Sellwood10}, but they 
still use two-dimensional approximation and the particle-mesh method with 
a grid size of $110 \times 128$ \citep{SB02}. Three-dimensional calculations
focused on the evolution of stellar bars 
\citep{Athanassoula05,DBS09,SD09}. Thus, it is important to 
investigate the basic physics of pure stellar disks using three-dimensional 
$N$-body simulations with a high enough resolution.

In this paper, we report the result of high-resolution $N$-body simulations 
of isolated stellar disks, in which multi-arm spirals spontaneously 
develop. We describe the method of our $N$-body simulations 
in Section 2. In Section 3, we show the results of simulations, and discuss 
the evolution of the spiral arms. We also discuss self-regulated mechanism of
spiral arms and how the maximum amplitude of spirals is determined.  
Section 4 is for summary and discussion.

\section{$N$-body simulations}

\begin{figure}[htbp]
\epsscale{1.0}
\plotone{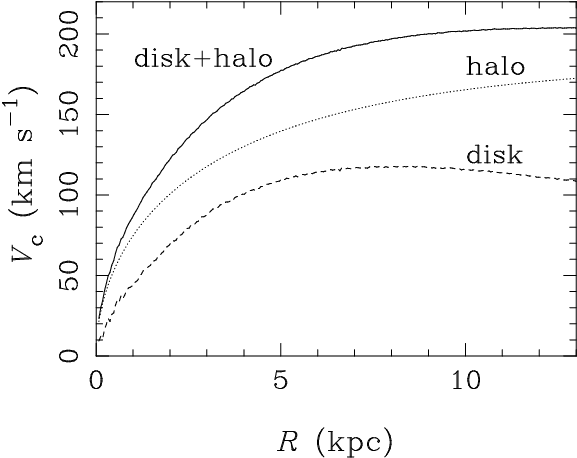}
    \caption{Circular velocity of model b (our standard model) as a
 function of the galactocentric distance.}
\label{fig:vc_b}
\end{figure}

We performed a series of $N$-body simulations of stellar disks in a 
fixed spherical dark halo potential. 
Our stellar disk and halo models are based on those in \cite{Baba09}.
Here, we briefly summarize the parameters and how we generate the 
initial equilibrium disks.
We adopted the NFW model \citep{NFW97} as the dark halo model
with the concentration parameter of the halo, $c=10$. 
The virial radius of halo, $R_{\rm h}$, is 122 kpc, and the mass 
within $R_{\rm h}$, $M_{\rm h}$ is $6.4 \times 10^{11}M_{\odot}$. 
{\footnote {
We assume that a spherical region, where the mean density is 200 times as 
high as a background density, is virialized. As the background cosmology, 
we adopted a concordance cold dark matter ($\Lambda$CDM) model with 
parameters: $\Omega_{\rm M}=0.3$, $\Omega_{\Lambda}=0.7$, and $H_0 = 70$km s$^{-1}$ 
Mpc$^{-1}$. The formation redshift of the halo is set to be 1.0.}} 
We adopted an exponential disk model as 
disk models. We varied total disk mass, $M_{\rm d}$, and initial $Q$ at
the reference radius (8.6 kpc in our models), $Q_0$. The scale radius, 
$R_{\rm d}$, is 3.4 kpc, and the scale height, $z_{\rm d}$, is 0.34 kpc.
We performed simulations with four different resolutions, 
$N=3\times 10^7, 9\times 10^6$, $3\times 10^6$, $1\times 10^6$, and 
$3\times 10^5$
(hereafter, 30M, 9M, 3M, 1M, and 300k, respectively).
We summarized our disk models in Table \ref{tb:model}. 
Hereafter, we regard model b ($M_{\rm d}/M_{\rm h}=0.050$ and $Q_0=1.2$) as our
standard model. The circular velocity of the disk at $R=10$ kpc is about
$200\,  \kms$ for model b. Its circular velocity profile is shown 
in Figure \ref{fig:vc_b}.
We generated initial disk models using the Hernquist method
\citep{H93}. Initially, the generated models are not exactly
in an equilibrium at the galactic center. Thus, they cause ripples
spreading through the disk from the center. To remove the ripples,
we integrated the models for a few Gyrs randomizing azimuthal 
positions of particles every four steps to prevent the growth of 
spiral arms \citep{MD07}.
After the ripples passed through the disk, we used it  
as the initial condition. 

We used a Burnes-Hut treecode \citep{BH86,M04} on GRAPE-7 \citep{Kawai06}
and GRAPE-DR \citep{Makino07}. The opening angle,
$\theta$, is 0.4 with the center-of-mass approximation. The maximum group
size for a GRAPE calculation \citep{M91}, $n_{\rm crit}$, is 2048.
For the time integration, we used a leapfrog integrator with a fixed
stepsize of $\Delta t=0.29$ Myr for $N=30$M, 9M, and 1M models and 
$\Delta t=0.15$ Myr for the other models. The gravitational potential 
is softened using the usual 
Plummer softening, with the softening length of $\epsilon = 30$ pc. This
is small enough to resolve the typical spiral structures ($\sim 100$ pc).
For all runs, the energy error was less than 0.1\% throughout
the simulations.
One may concerned about the drift of the disk in the halo potential because 
treecodes do not conserve the linear momentum of the system. Therefore,
we investigated the position of the density center of the disk and 
confirmed that the drift does not occur (for the details, see Appendices 
A and B). 

\begin{table}[htbp]
\caption{Initial disk models}
\begin{center}
\begin{tabular}{c|ccc}
\hline
Model    & $Q_0$   & $M_{\rm d}/M_{\rm h}$ & $N$\\ \hline \hline
a & 1.1 & 0.050 & 3M/300k \\
b & 1.2 & 0.050 & 30M/9M/3M/1M/300k\\
c & 1.3 & 0.050 & 3M/300k\\
d & 1.4 & 0.050 & 3M/300k\\
e & 1.5 & 0.050 & 3M/300k\\
f & 1.8 & 0.050 & 3M/300k \\ \hline
g & 0.5 & 0.050 & 3M\\
h & 1.2 & 0.075 & 3M\\
i & 1.2 & 0.033 & 3M\\
\hline
\end{tabular}
\end{center}
\label{tb:model}
\end{table}

\section{Evolution of stellar disks}

\subsection{Evolution of spiral arms}
\begin{figure*}[htbp]
\epsscale{1.0}
\plotone{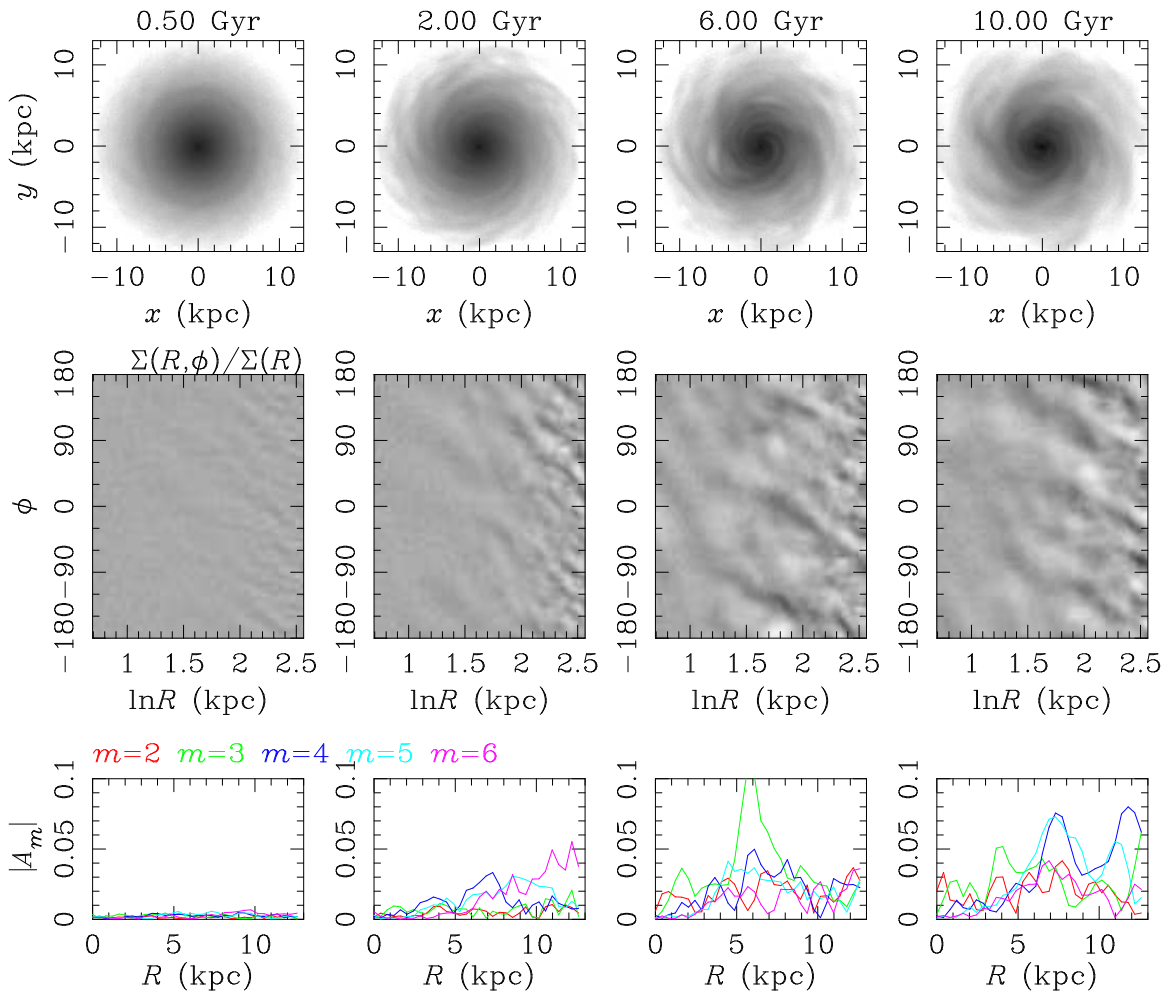}
   \caption{Evolution of spiral arms for model b with $N=30$M. 
Top panels show the surface
 density, middle panels show the surface density normalized at each
 radius, and bottom panels show the Fourier amplitudes.}
\label{fig:snapshot1_30M} 
\end{figure*}

\begin{figure*}[htbp]
\epsscale{1.0}
\plotone{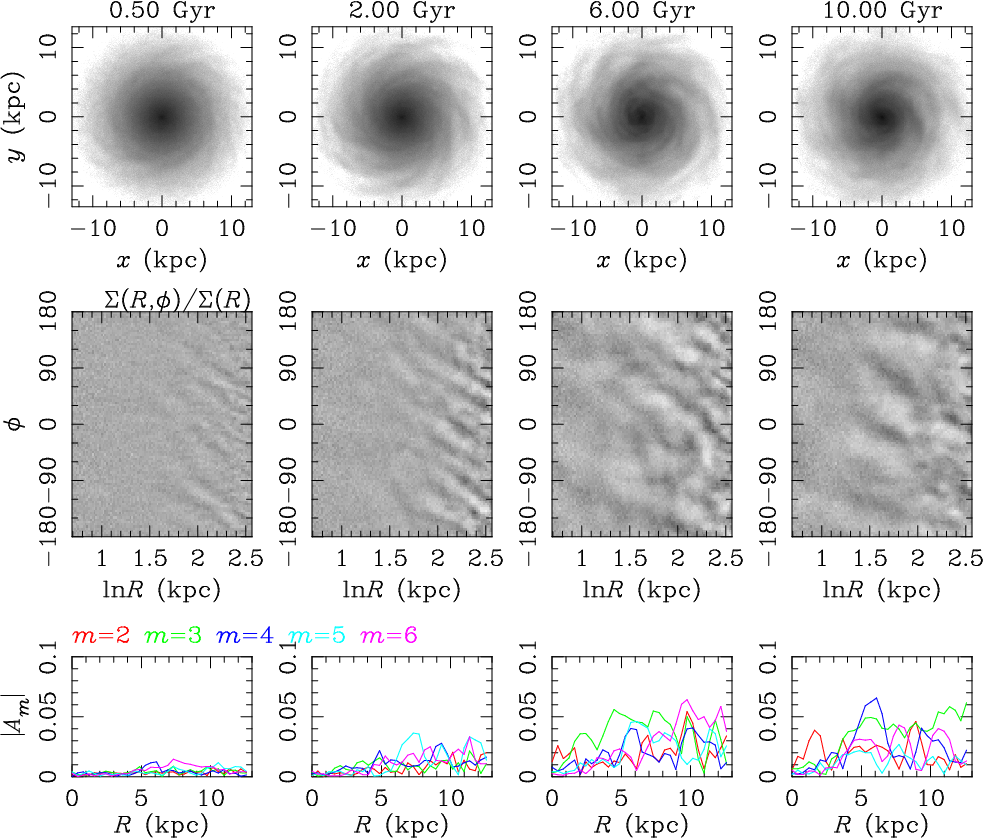}
   \caption{Same as Figure \ref{fig:snapshot1_30M}, but for $N=9$M.}
\label{fig:snapshot1_9M} 
\end{figure*}

\begin{figure*}[htbp]
\epsscale{1.0}
\plotone{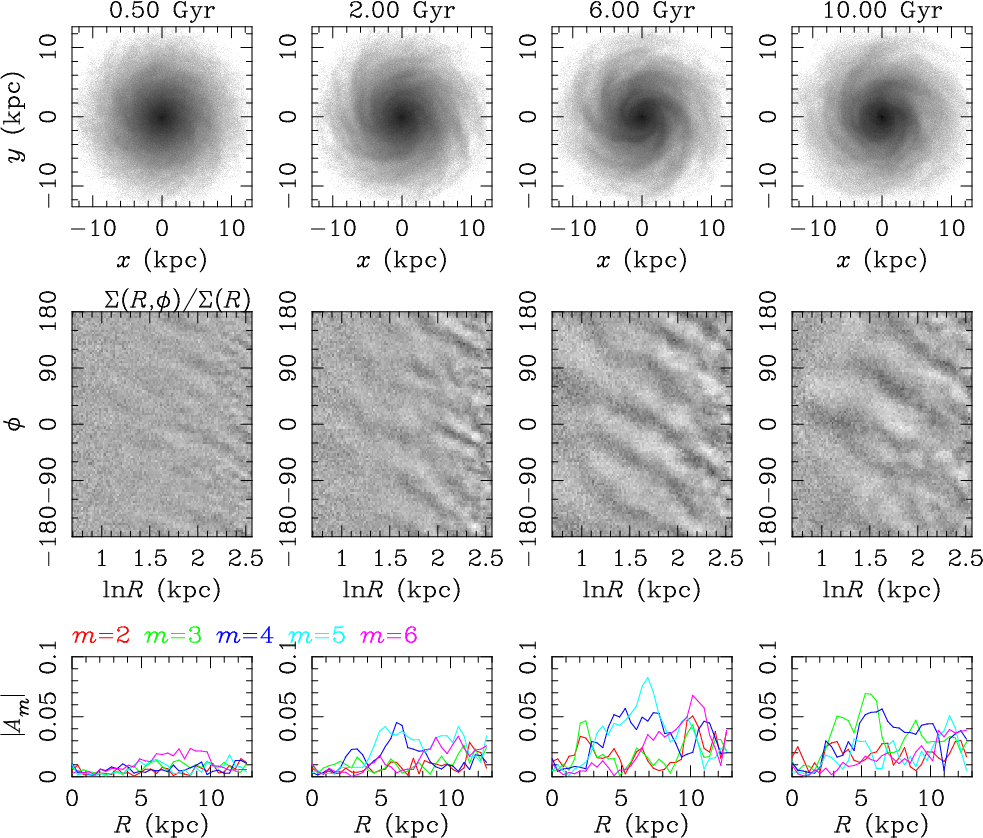}
   \caption{Same as Figure \ref{fig:snapshot1_30M}, but for $N=3$M.}
\label{fig:snapshot1} 
\end{figure*}

\begin{figure*}[htbp]
\epsscale{1.0}
\plotone{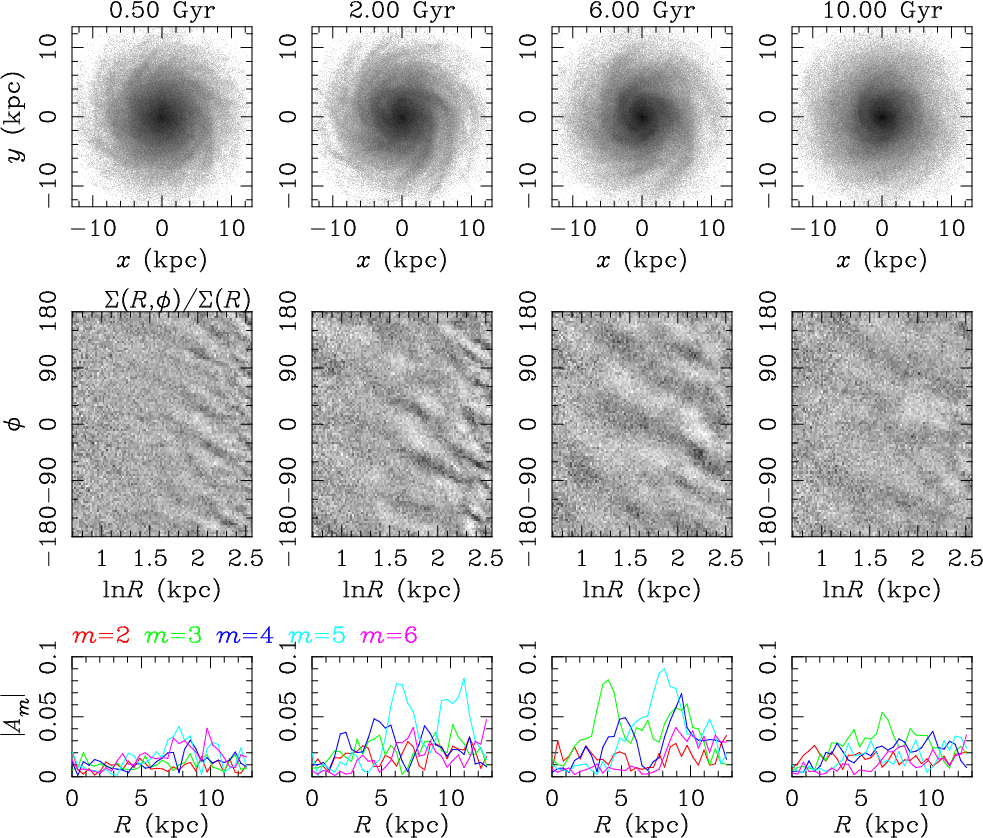}
   \caption{Same as Figure \ref{fig:snapshot1_30M}, but for $N=1$M.}
\label{fig:snapshot1_1M} 
\end{figure*}

\begin{figure*}[htbp]
\epsscale{1.0}
\plotone{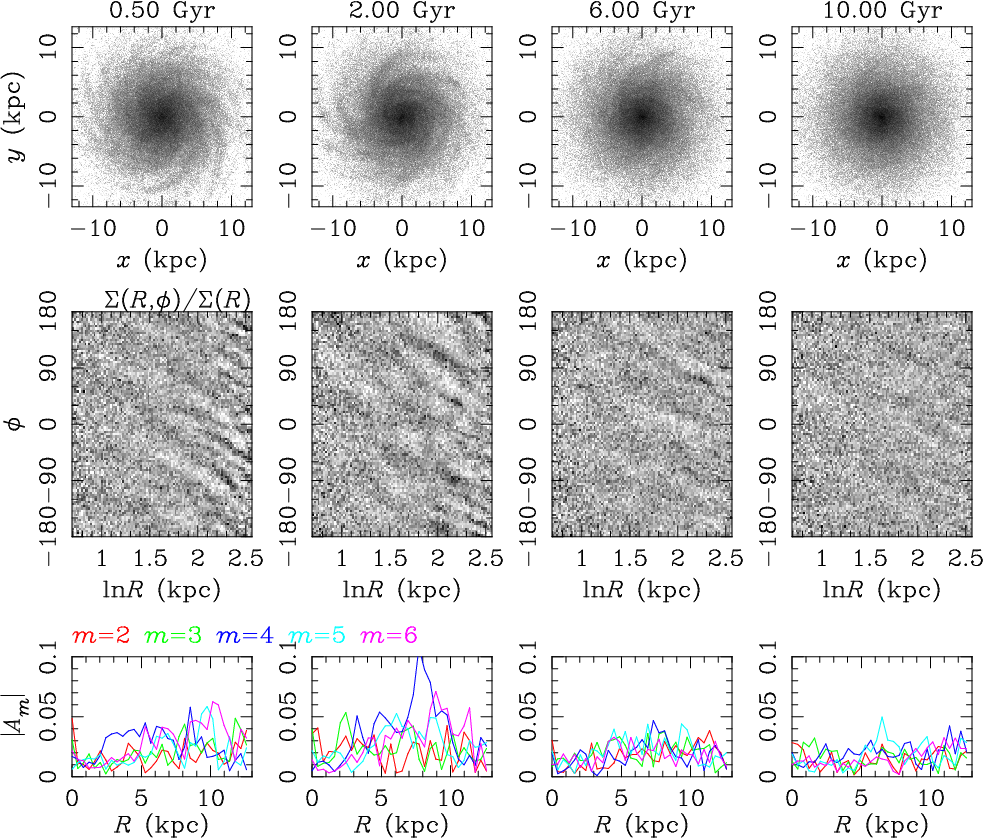}
    \caption{Same as Figure \ref{fig:snapshot1_30M}, but for $N=300$k.}
 \label{fig:snapshot1_300k}
\end{figure*}

First, we show the evolution of spiral arms of our standard model, model
b.  Figures \ref{fig:snapshot1_30M} --
\ref{fig:snapshot1_300k} show the evolution of model b for $N=30$M,
9M, 3M, 1M, and 300k, respectively. Top panels show the surface
density of the disk in the Cartesian coordinate.  Middle panels
show the density contrast, $\Sigma (R, \phi)/\Sigma (R)$, in the polar
coordinate. Here, $\Sigma (R, \phi)$ and $\Sigma (R)$ are averaged 
surface density
in a polar grid from $R$ to $R+\Delta R$ and from $\phi$ to 
$\phi + \Delta \phi$ and a ring from $R$ to $R + \Delta R$, respectively. 
We used $\Delta R = 1$ kpc and $\Delta \phi =\pi/64$.  
Bottom panels show Fourier amplitudes at each radius obtained from a 
Fourier series:
\begin{eqnarray}
 \frac{\Sigma (R,\phi)}{\Sigma (R)} = \sum^{\infty}_{m=1} A_m \exp 
[-im\phi],
\label{eq:Fourier}
\end{eqnarray}
where $m$ is the azimuthal wavenumber (i.e., the number of spiral
arms) and $A_m$ is the Fourier amplitude. 
Only the amplitudes of $m=$2--6 are shown 
in this figure. Other modes are much smaller than these modes throughout 
the simulations.

\begin{figure*}[htbp]
\epsscale{1.0}
\plottwo{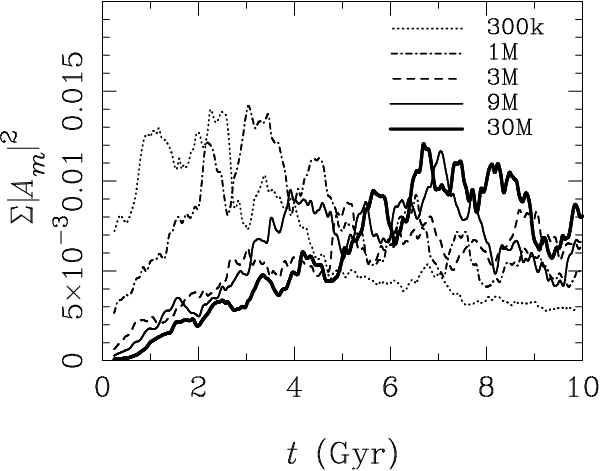}{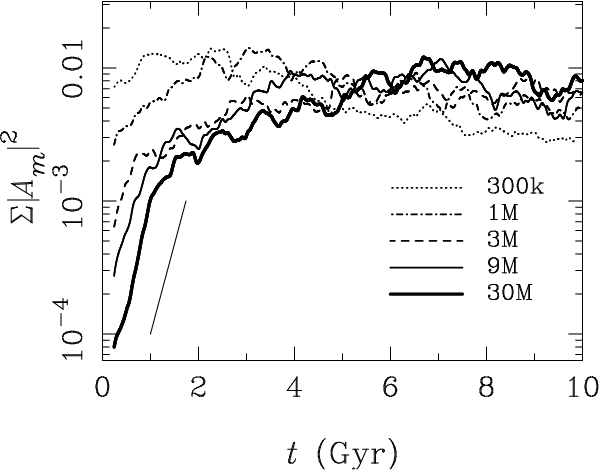}
   \caption{Left panel: time evolution of total power 
($\sum^{10}_{m=1}|A_m|^2$) for model b at $R=7.5$ kpc. Right panel: same as
the left panel, but in the logarithmic scale. The solid thin line shows an 
exponential timescale of 0.37 Gyr.\label{fig:tp_b}}
\end{figure*}

Initially many spiral arms with small amplitudes appear
(see $t=0.50$ Gyr in Figures \ref{fig:snapshot1_30M} -- 
\ref{fig:snapshot1_300k}). Eventually they merge and 
the amplitude of small wavenumber such as $m=4$ becomes larger. 
Figure \ref{fig:tp_b} shows the total power, the sum of 
squared amplitudes defined as $\sum _{m=1}^{10}|A_m|^2$, at 
$7.5\pm0.5$ kpc. The data points are averaged over 0.5 Gyr.
In the beginning of the simulations, the total powers
grow exponentially from their initial amplitudes, which are determined
by Poisson noise. The growth timescale is $\sim 0.4$Gyr. This value is 
comparable to that in the case of the bar mode \citep{DBS09}.
In the case of $N=30$M, we performed the simulation only to 5 Gyr.
However, the evolution in the run is quite similar to those 
in the runs with $N=3$M, 9M, and 30M, although the total power grows from a 
smaller value.
The total powers reach their peak values at 
$t\sim 2$ Gyr for $N=300$k and $t\sim 3$ Gyr for $N=1$M. In the case
of $N=3$M, 9M, and 30M, the peak is not clear, but the total powers have
developed well at $t\gtrsim 6$ Gyr. Clearly, the dependence
on the number of particles exists. In the case of a larger number of
particles, it takes more time for the spiral arms to develop because they 
start from a smaller amplitude of Poisson noise. 
After the spiral arms have developed, their number is 
consistent with that expected from the swing amplification theory 
\citep{Toomre81}, in which spiral arms with $1<k_{\rm cr}R/m<2$, where 
develop most efficiently. In our model b, $m\simeq4$ for $k_{\rm cr}R/m=1.5$.

After $t \sim 3$ Gyr, the amplitudes of $N=300$k model start to decay
(see Figure \ref{fig:tp_b}), and the
spiral arms almost disappear at $t=6$ Gyr 
(see Figure \ref{fig:snapshot1_300k}). The behavior of 1M model is
similar to the 300k model.
On the other hand, the spiral arms in $N=$3M, 9M, and 30M models are still 
prominent after 6 Gyr, and their amplitudes are $|A_m| \sim 0.05$
even at the end of 
the simulation, $t=$ 10 Gyr, in contrast to $|A_m| \lesssim 0.03$ in 
$N=300$k model (see Figure \ref{fig:snapshot1})

Evolutions of the amplitudes of spiral arms are qualitatively different
between higher and low resolutions. In $N=300$k model, the amplitude 
grows rapidly 
in the first 2 Gyr, and after this rapid growth it decreases fairly
rapidly. On the other hand, the amplitude in $N=3$M, 9M, and 30M models 
keeps growing to the end of simulation, i.e., $t=10$ Gyr. 
The highest resolution run may show the same evolution in $>5$ Gyr.

We found that spiral arms developed in the disk are not stationary, but 
timedependent. As shown in Figure \ref{fig:tp_b}, the amplitude of the
arms in $N=3$M and 9M models at $R = 7.5 $ kpc oscillates 
quasi-periodically in the timescale of $\sim 1$ Gyr.
Figures \ref{fig:snapshot1_30M} -- \ref{fig:snapshot1_300k} show that all 
modes of spiral arms are 
timedependent and the dominant mode changes spatially.
Each spiral arm is wound up because of the galactic differential 
rotation. As a result, the global spiral arms break up into smaller 
fragments whose sizes are typically a few kpcs. These fragments eventually 
collide and reconnect with other fragments
due to the differential rotation, and the global spiral arms revive.
This process of breaking up and reconnection repeats throughout the 
simulations.

\begin{figure*}[htbp]
\epsscale{1.00}
\plottwo{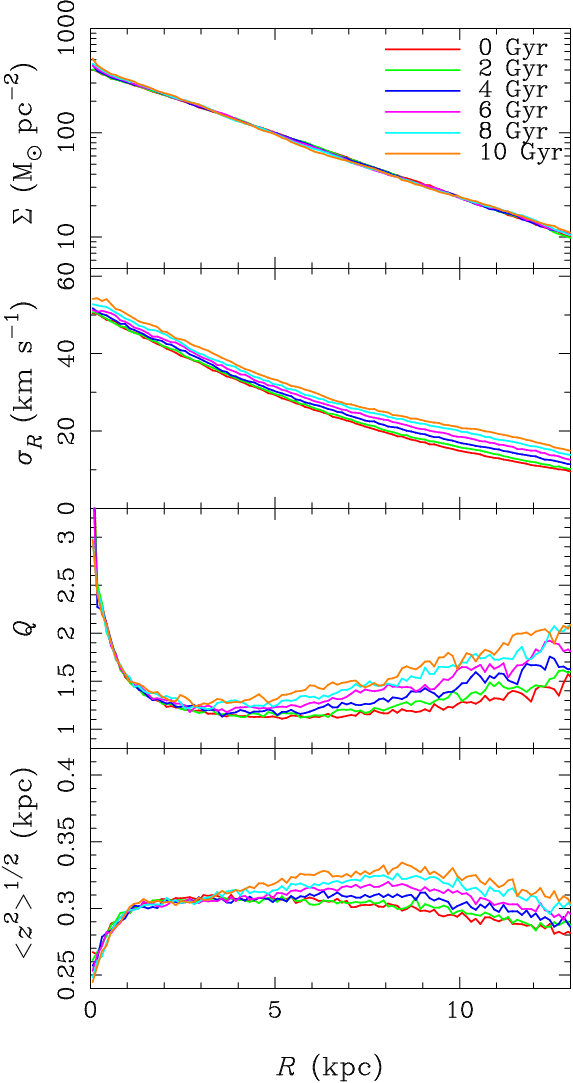}{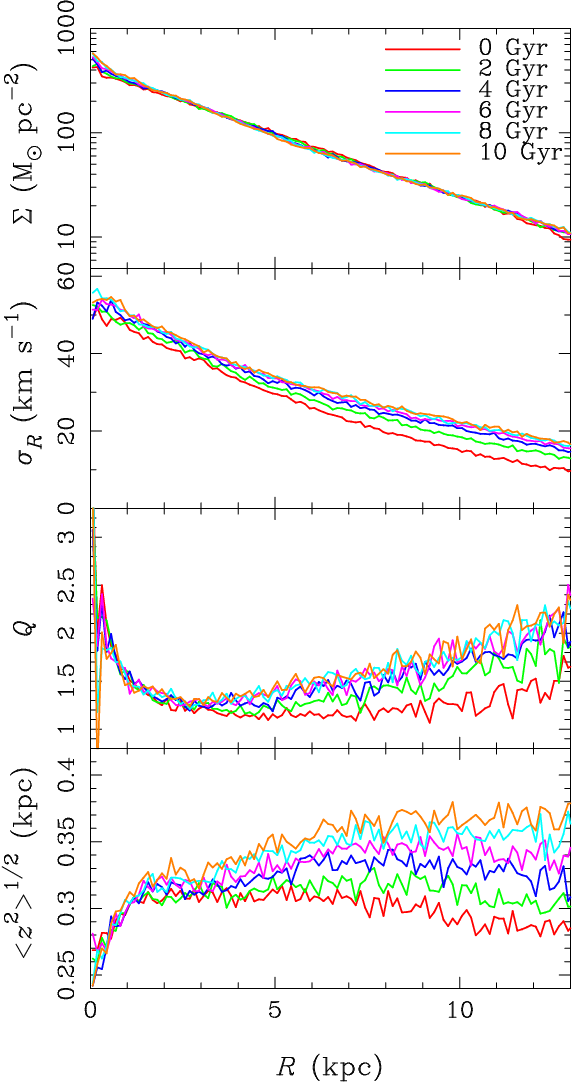}
\caption{Evolution of surface density, $\Sigma$, radial velocity
 dispersion, $\sigma_R$, Toomre's $Q$ value, and scale height, 
$\langle z^2\rangle ^{1/2}$, of disk for Model a1, $N=3$M (left) and 300k 
(right). Here, we defined the rms of the vertical position of
 stars as $\langle z^2\rangle^{1/2}$.}
\label{fig:ev_b}
\end{figure*}

Figure \ref{fig:ev_b} shows the radial distribution of 
the surface density, $\Sigma$, radial velocity dispersion, $\sigma_R$, 
Toomre's $Q$, and scale height, $\langle z^2\rangle ^{1/2}$, of model 
b for $N=3$M and $300$k. These values are averaged over bins of 500 pc.
While the distribution of $\sigma_R$ and $Q$ at $t=10$ Gyr are very 
similar in both models, $\langle z^2\rangle ^{1/2}$ in $N=300$k increases more 
rapidly than that in $N=3$M. 
The evolution of $\langle z^2 \rangle ^{1/2}$ is 
caused by the two-body relaxation. The two-body relaxation time
of this model is $\sim 3$ Gyr and $\sim 30$ Gyr for $N=300$k and 3M.
The evolutions of $\sigma_R$ and $Q$ are 
also faster in $N=300$k models.
We will discuss the effects of the number of particles in Section 3.5.

\subsection{Evolution of the $Q$ value due to the spiral heating}

\begin{figure*}[htbp]
\epsscale{1.0}
\plottwo{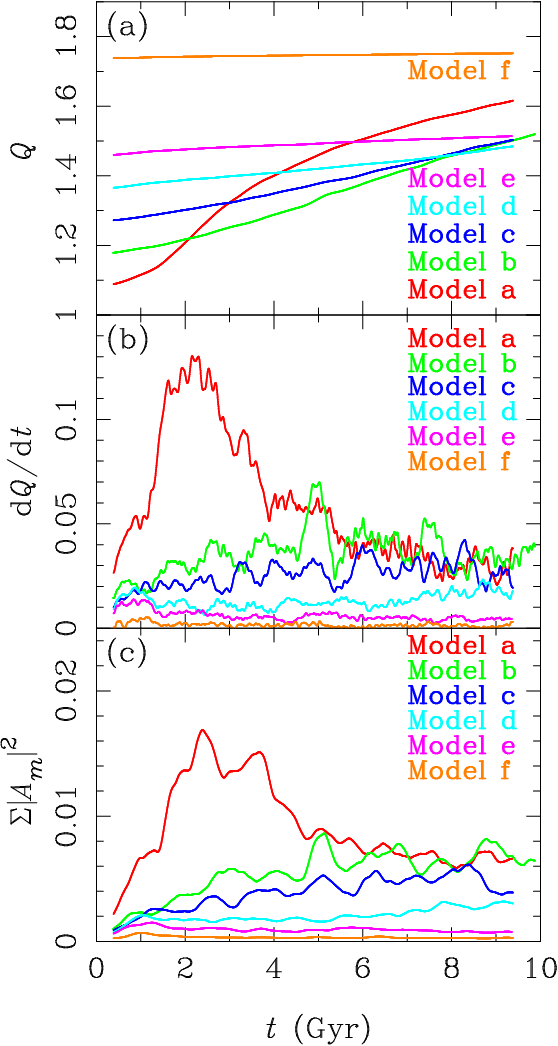}{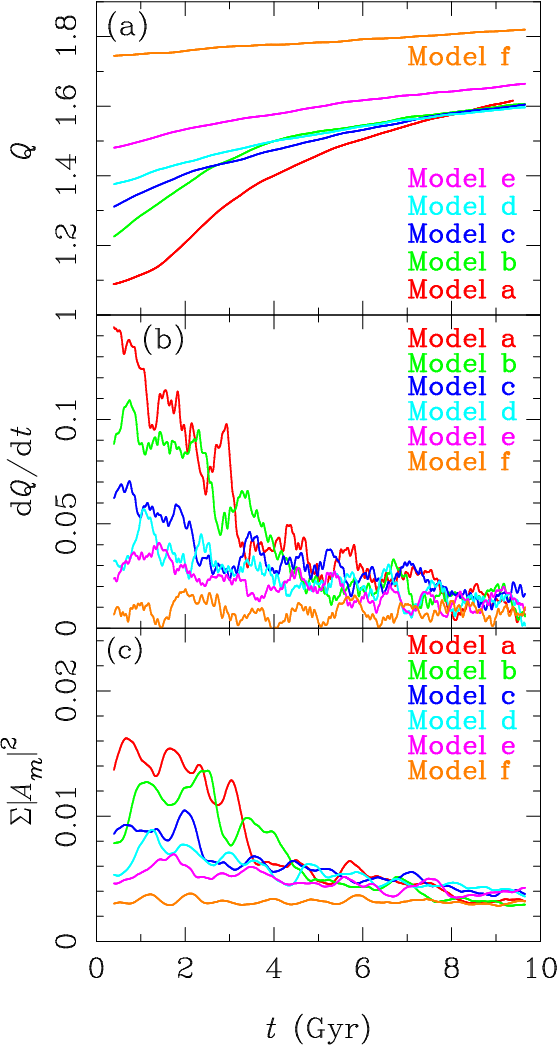}
\caption{Time evolution of $Q$ (a), $dQ/dt$ (b), and total power
 (c) for model b, with $N=3$M (left) and 300k (right). These $Q$ 
and $dQ/dt$ are averaged 
between 5--10 kpc, and the total powers are evaluated at $7.5\pm 0.5$
kpc.}
 \label{fig:three}
\end{figure*}

In this section, we compare the results of models with different 
initial values of $Q$, $Q_0$, and investigate how the $Q$ evolve 
in time. We performed simulations for models with 
different $Q_0$ (model a--f). 
Top panels of Figure \ref{fig:three} show the time (a)
evolution of $Q$, (b) the growth rate of $Q$, i.e., $dQ/dt$, 
and (c) the total power of the modes, $\sum |A_m|^2$, which corresponds 
to the amplitude of spiral arms.  These values of $Q$ and $dQ/dt$ are 
averaged over the range of 5--10 kpc and over 0.5 Gyr. The total powers 
are evaluated at $7.5 \pm 0.5$ kpc.
If the initial disk is colder (i.e., smaller $Q_0$), the amplitudes 
of spiral arms tend to be larger in both $N=300$k and 3M models.
It is also clearly seen that $Q$ increases more rapidly in colder
initial disks (i.e., smaller $Q_0$).
In all $N=3$M models, the amplitudes of spirals tend to increase
toward $t=10$ Gyr, except for model a.
On the other hand, the amplitudes start to decrease soon after the
simulations start in all $N=300$k models (right bottom panel of Figure 
\ref{fig:three}).
The peak amplitude is larger for models with smaller $Q_0$, for
both numbers of particles.

From comparison between panels (a) and (c) of Figure 
\ref{fig:three}, it seems that the increase of 
the $Q$ is caused by the spiral arms, since the $Q$ 
rapidly increases when the amplitude of spiral arms is large. 
Since the surface density, $\Sigma$, and epicycle frequency, $\kappa$,
do not change significantly throughout the simulations (see Figures 
\ref{fig:ev_b}), the change of $Q$ 
depends only on the radial velocity dispersion, $\sigma_R$, by the 
definition of $Q$:
\begin{eqnarray}
Q = \frac{\sigma_{R}\kappa}{3.36G\Sigma}.
\label{eq:Q}
\end{eqnarray}
Therefore, $dQ/dt$ can be interpreted as  a ``heating rate.''
The evolution of $dQ/dt$ and that of the total power of the 
spiral arms, $\sum |A_m|^2$, are very similar (see the left panels 
(b) and (c) of Figure \ref{fig:three}), indicating 
that the spiral arms increase the velocity dispersion of stars.
The similarity between $dQ/dt$ and $\sum |A_m|^2$ is also visible 
in the 300k models (right panels (b) and (c) of Figure \ref{fig:three}). 
Therefore, this mechanism seems to be 
independent of the number of particles.

\begin{figure*}[htbp]
\epsscale{1.00}
\plottwo{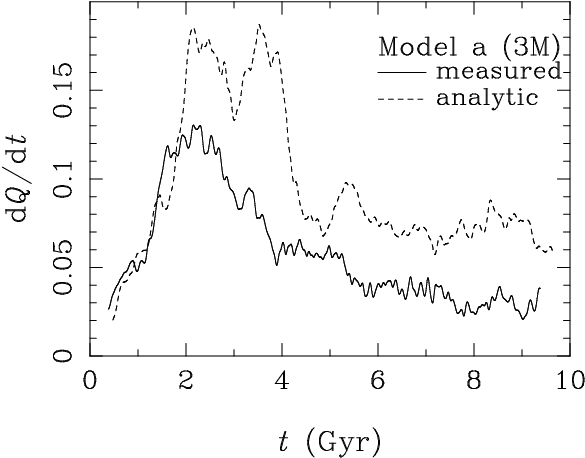}{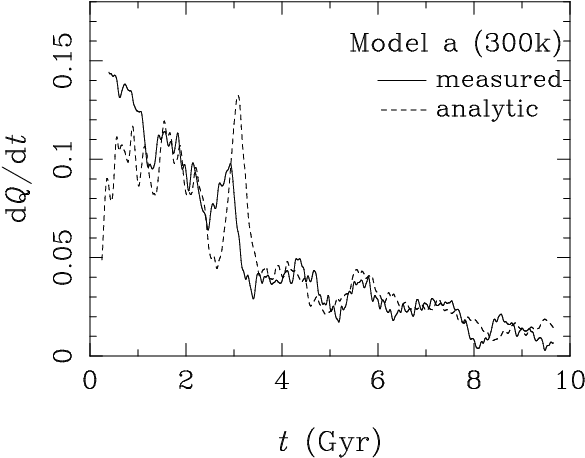}
 \caption{Comparison of $dQ/dt$ between that obtained from simulations 
(measured) and that estimated from Equation (\ref{eq:dQdt2}) (analytic) 
 for model a, $N=3$M (left) and $N=300$k (right) at $R=7.5$ kpc.
\label{fig:comp_a}}
\end{figure*}

\begin{figure*}[htbp]
\epsscale{1.00}
\plottwo{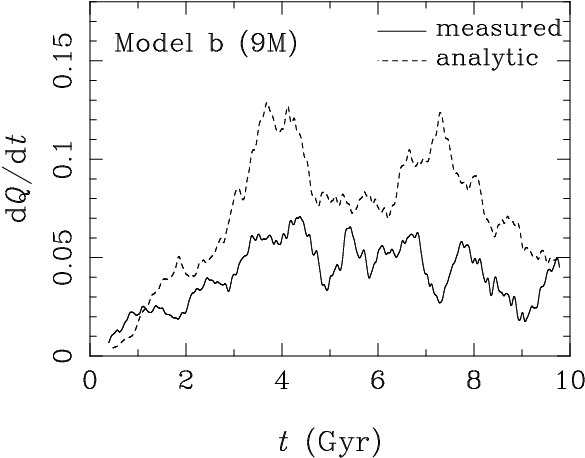}{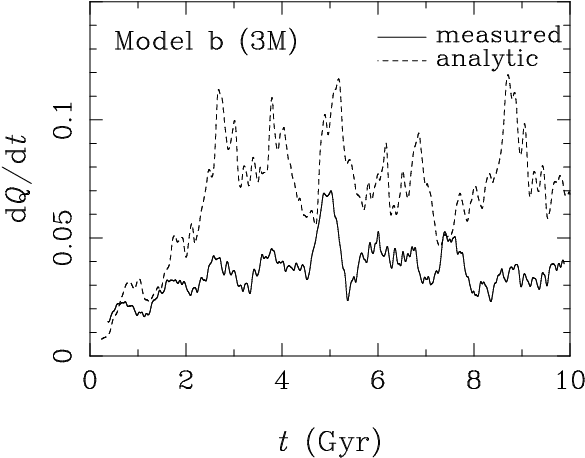}
\plottwo{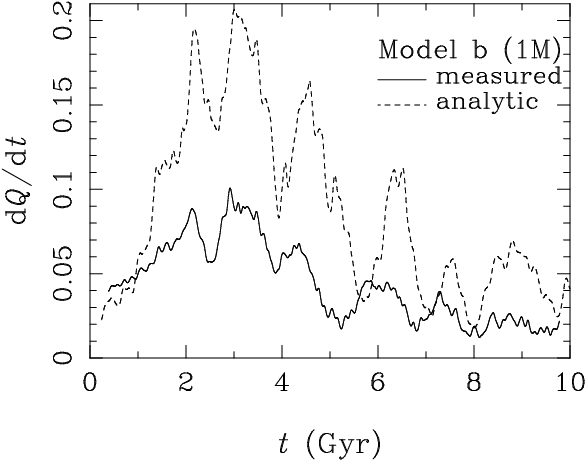}{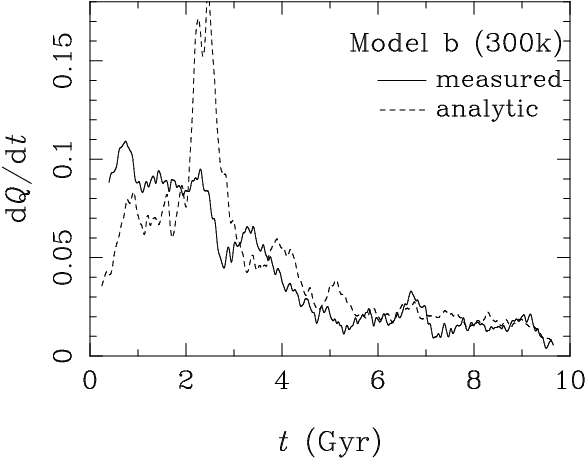}
 \caption{Same as Figure \ref{fig:comp_a}, but for model b.
 \label{fig:comp_b}}
\end{figure*}

In order to confirm the hypothesis that the amplitude of spiral arms, 
$\sum |A_m|^2$, determines the heating rate, $dQ/dt$, we analytically 
estimate $dQ/dt$ from the amplitude of spiral arms in the simulations. 
The relation that $dQ/dt$ is proportional to
the square of amplitudes is suggested by \citet{CS85}. They derived this 
relation by considering the perturbing potential of spiral arms. We 
derive the relation between $dQ/dt$ and the spiral amplitude in a
different way.
As shown in the previous section, the global spiral arms are transient; 
splitting to smaller sub-arms and merging into global arms recurrently 
occur. Therefore, as a first approximation, they behave like 
`material arms' consisting of several massive clumps. With this 
assumption, we can estimate the increase of stellar velocity dispersion
using the same equation as that describes dynamical heating of stars by 
GMCs in galactic disks \citep{KI92}, replacing a 
spiral arm in our simulation with several massive clumps. 
As discussed above, the evolution of $Q$ corresponds to that of 
the velocity dispersion. We therefore 
discuss the time derivative of the velocity dispersion below.
The relaxation time of disk stars due to the dynamical heating by 
clumps with mass $M_{\rm c}$, whose number density is $n_{\rm c}$, is given by
\begin{eqnarray}
t_{\rm g} \simeq \frac{v^3}{\pi n_{\rm c} G^2 (M_{\rm c}+m_{\rm s})^2\ln \Lambda},
\end{eqnarray}
where 
$v$ and $m_{\rm s}$ are the three-dimensional velocity dispersion of disk stars 
and mass of a star respectively, and $\ln \Lambda$ is Coulomb logarithm
\citep{BT08}. Here, since $M_{\rm c} \gg m_{\rm s}$, we can neglect $m_{\rm s}$. 
The number density of stars in spiral arms is given by 
$n_{\rm c} = \Sigma_{\rm c}/(\langle z^2\rangle ^{1/2}M_{\rm c})$, where 
$\langle z^2 \rangle ^{1/2}$ is the scale height of the disk and $\Sigma_{\rm c}$ 
is the surface density of the spiral arms. 
The scale height $\langle z^2\rangle ^{1/2}$ is given by 
$\langle z^2\rangle ^{1/2} \simeq \sigma _z/\Omega$, where $\sigma _z$ is the 
vertical velocity dispersion and $\Omega$ is the angular speed of the disk. 
In a disk system, $\sigma _z \simeq v$ \citep[e.g.,][]{KI92}, 
the relaxation time of a disk can be written as
\begin{eqnarray}
  t_{\rm g} \simeq \frac{v^4}{\pi \Sigma_{\rm c} G^2 M_{\rm c} \Omega \ln \Lambda}.
\label{eq:tg}
\end{eqnarray} 
By definition, the relaxation time is 
$t_{\rm g} = \frac{v^2}{{dv^2}/{dt}}$.
The change of velocity dispersion is written as
\begin{eqnarray}
  \frac{dv}{dt} = \frac{1}{2v}\frac{dv^2}{dt} = \frac{v}{2t_{\rm g}}.
  \label{eq:dvdt}
\end{eqnarray}
We assume that the mass and surface density of each clump of the spiral 
arms are given by 
\begin{eqnarray}
  M_{\rm c} &=& M_{\rm d} \sum^m \frac{|A_m|}{m^2},\label{eq:a}\\
  \Sigma_{\rm c} &=& \Sigma \sum^m |A_m|,\label{eq:b}
\end{eqnarray}
where $M_{\rm d}$ and $\Sigma$ are the total mass and surface density of the 
disk, and $m$ and $A_m$ are the number and amplitude of the spiral arms. 
Substituting Equations (\ref{eq:tg}), (\ref{eq:a}), and (\ref{eq:b}) to 
Equation (\ref{eq:dvdt}), we obtain
\begin{eqnarray}
 \frac{dv}{dt} \simeq \frac{\pi \Sigma G^2 M_{\rm d} 
\Omega \ln \Lambda}{2v^3} \sum^m \frac{|A_m|^2}{m^2}.
 \label{eq:dvdt2}
\end{eqnarray}
From Equation (\ref{eq:Q}), the time derivative of $Q$ is
\begin{eqnarray}
  \frac{dQ}{dt} \simeq \frac{\kappa}{3.36 G \Sigma} \frac{d\sigma_R}{dt},
  \label{eq:dQdt}
\end{eqnarray}
where we assumed that $\kappa$ and $\Sigma$ are constants. Assuming
that the three-dimensional velocity dispersion is $v=\sqrt{3}\sigma_R$, 
from 
Equations (\ref{eq:dvdt2}) and (\ref{eq:dQdt}), we obtain
\begin{eqnarray}
 \frac{dQ}{dt}  \simeq \frac{\pi \kappa G M_{\rm d} \Omega \ln \Lambda}{11.6 v^3} 
		\sum^m \frac{|A_m|^2}{m^2}.
\end{eqnarray}
If we scale as $R=8 ({\rm kpc})=1$, $M_{\rm d}=3.2\times 10^{10} M_\odot=1$, 
and $\Omega = 190 \kms =1$,
then
\begin{eqnarray}                  
 \frac{dQ}{dt} \simeq 3.9 \ln \Lambda \left( \frac{190\ [\kms]}{v} \right) ^3 
\sum^m
  \frac{|A_m|^2}{m^2} \ \mathrm{Gyr^{-1}},
 \label{eq:dQdt2}
\end{eqnarray}
where $G=0.39$ and $\kappa\simeq 1.5$ in our model.
Using Equation (\ref{eq:dQdt2}) and the amplitudes, $|A_m|^2$, obtained 
from the simulation, we can estimate $dQ/dt$. 
We take the sum of Fourier components with $m=$4--6, which are dominant
modes. As can be seen in the snapshots and maps of 
$\Sigma (\phi,R)/\Sigma (R)$ (Figures 
\ref{fig:snapshot1_30M}--\ref{fig:snapshot1_300k}),  
it would be unphysical to include modes with $m\le 3$.
We adopted $\ln \Lambda = 1.0$ because the scale height of the disk is 
comparable to the size of the clumps.
Figures \ref{fig:comp_a} and \ref{fig:comp_b} 
show the comparison between the analytic and the numerical results of 
model a and b.  It is clear that behavior of $dQ/dt$ in the 
simulations is quantitatively reproduced by the analytic estimate in 
models a and b, for both $N=3$M and 300k.
Thus, we conclude that scattering of stars by spiral arms can 
heat up the stellar disks, and that the heating rate is 
proportional to the squared amplitude of the spiral arms.

\subsection{The maximum amplitude of spiral arms}
In the previous section, we showed that  $dQ/dt$ is tightly coupled 
with the amplitude of spiral arms, and this coupling is well understood
in a simple picture that spiral arms gravitationally scatter stars and 
increase $Q$.
What controls the amplitude of spiral arms? Figure \ref{fig:three} 
shows that the amplitudes of spiral arms decay, 
as $Q$ increases. This suggests that the amplitude of spiral arms 
is determined by $Q$.

\begin{figure}[htbp]
\epsscale{1.0}
\plotone{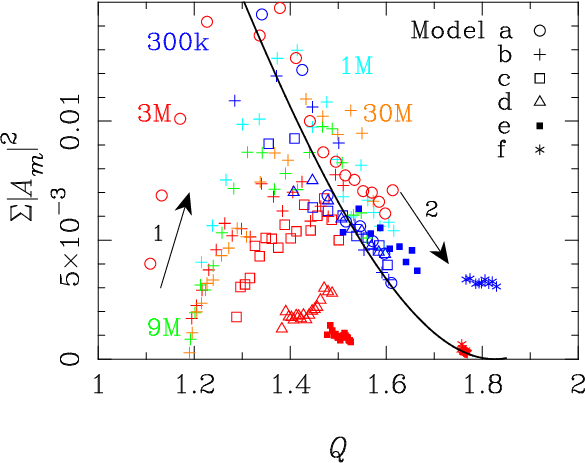}
\caption{Relation between $Q$ and total power of spiral amplitude. 
The value is evaluated at 7.5 kpc and averaged in 0.5 Gyr.
Black curve shows 
$(0.1A_m)^2$ obtained from equation (\ref{eq:amp2}).
Arrows shows the direction of the time evolution.\label{fig:Q_tp}}
\end{figure}

Figure \ref{fig:Q_tp} shows the evolution of models in the plane of $Q$ 
and the total power of the spiral arms. These values are measured 
at 7.5 kpc and averaged over 0.5 Gyr in the same way as the previous 
section. At the beginning of the simulation, 
both $Q$ and the total power are small.
The amplitude of spiral arms grows rapidly, and the models move in the 
right-upward direction as shown by arrow 1. Once the amplitudes
reach their peak values, they decay and models move
right-downward (arrow 2). In this phase, the trajectory in the 
$Q$-$\sum |A_m|^2$ plane seems to follow a roughly straight line,
irrespective of models and the number of particles. In other words, 
there seems to be a ``forbidden region'' in the left-top of the 
$Q$-$\sum |A_m|^2$ plane, where both amplitudes and $Q$ 
are large. This result implies that the maximum amplitude is determined  
by $Q$. At the beginning, the amplitude is smaller
than the maximum amplitude and therefore the spiral arms can grow
with time. However, once the amplitude reaches to its limit, 
it starts to decay because $Q$ increases due to heating 
by spiral arms, and the maximum amplitude decreases.

Here we try to estimate the maximum amplitude of spiral arms,
assuming that the spiral arms grow through the collapse due to the 
gravitational instability, and that they evolve until they reach an 
approximate dynamical equilibrium.
Under this assumption, the amplitudes can be simply estimated as a 
density contrast before and after the collapse. We assume that 
stars in a region of the disk with a size of a critical wavelength, 
$\lambda_{\rm cr}$, collapse to a spiral arm.

The initial energy of the region to collapse can be expressed as
\begin{eqnarray} 
E_0 = K_0 + W_0 = \frac{1}{2}M\sigma ^2 - C\frac{GM^2}{r_0},
\label{eq:energy}
\end{eqnarray}
where $M$, $\sigma$, and $r_0$ are the mass, velocity dispersion, and
radius of the system, and $C$ is a fixed value. We treat $C$ as a parameter,
depending on geometry and density distribution; e.g., for a 
homogeneous sphere, $C=3/5$.
Assuming that the collapsed region is virialized when it forms
a spiral arm, the arm satisfies the virial theorem and the potential 
energy after the virialization is 
\begin{eqnarray}
W=2E_0=2(K_0+W_0).
\label{eq:virial}
\end{eqnarray}
The amplitude of the spiral arms is obtained
from the density contrast of the initial and virialized density.
If we assume that the virialized density is the mean density 
inside the half-mass radius, $r_{\rm h}$, 
we obtain the amplitude from the ratio of the initial and 
virialized densities,
\begin{eqnarray}
\frac{\rho}{\rho _0} = \frac{M/(2r_{\rm h}^3)}{M/r_0^3} 
 = \frac{1}{2}\frac{r_0^3}{(0.45GM^2)^3}|2(K_0+W_0)|^3,
\label{eq:rho}
\end{eqnarray}
where we adopt the half-mass radius, $r_{\rm h} = 0.45GM^2/|W|$ \citep{BT08}, 
and $W$ is obtained from Equation (\ref{eq:virial}).
From Equation (\ref{eq:Q}) and the critical wavelength, 
$\lambda _{\rm cr} = 4\pi^2G\Sigma/\kappa^2$, we obtain 
\begin{eqnarray}
\sigma _{R}^2 = \frac{(3.36)^2}{4\pi^2}G\Sigma \lambda_{\rm cr}Q^2.
\label{eq:sigma}
\end{eqnarray}
Assuming that $\sigma = \sqrt3 \sigma _R$ and the radius of the sphere, 
$r_0 = \lambda_{\rm cr}/2$, we can rewrite Equation (\ref{eq:rho}) using 
Equations (\ref{eq:energy}) and (\ref{eq:sigma}) as
\begin{eqnarray}
\frac{\rho}{\rho _0} = \frac{1}{2}\left(4.4C - 
		0.95\frac{\Sigma \lambda_{\rm cr}^2}{M}Q^2\right)^3,
\label{eq:rho2}
\end{eqnarray}
where we assumed that $E_0<0$.
Since $M\sim \Sigma \lambda_{\rm cr}^2$, the density contrast is written as a
function of $Q$,
\begin{eqnarray}
\frac{\rho}{\rho _0} = \frac{1}{2}\left(4.4C - 0.95Q^2\right)^3.
\label{eq:rho3}
\end{eqnarray} 
While the density contrast in Equation (\ref{eq:rho3}) is defined in
three dimensions, the amplitude we obtained from the simulation 
is one-dimensional because it is the contrast of the radially averaged 
surface density. Therefore, we define the amplitude of a spiral arm as
\begin{eqnarray}
A_m \equiv \left(\frac{\rho}{\rho_0}\right)^{1/3} - 1.
\label{eq:amp}
\end{eqnarray}
From Equations (\ref{eq:rho3}) and (\ref{eq:amp}), the amplitude relates with
$Q$ as
\begin{eqnarray}
A_m = 3.5C - 1.0 - 0.75Q^2.
\label{eq:amp2}
\end{eqnarray}
The black curve in Figure \ref{fig:Q_tp} shows $(0.1A_m)^2$ obtained from
Equation (\ref{eq:amp2}), where we assumed $C=1.0$. Although Equation 
(\ref{eq:amp2}) qualitatively explains the amplitude as a function of $Q$ 
after it reaches the maximum, 
the amplitude obtained from the simulations is smaller than that obtained 
from Equation (\ref{eq:amp2}) by a factor of 10. The possible 
reasons are as follows. 
(1) We assumed that a homogeneous region collapses to estimate the density 
contrast, but this is not the case in a disk. Especially, 
the scale height of the disk is much smaller than $r_0$, which we assumed
as the initial radius. (2) We used the radially averaged surface density to
calculate the Fourier amplitude from the simulations. This treatment may 
underestimate the local amplitude of spiral arms because the averaged 
density depends on the 
radial width for averaging. We chose a radial width smaller than the 
critical wavelength. However, the amplitudes increased by $\sim 10$\% 
when we halved the width. (3) Growth of spiral arms does not complete 
because the galactic shear breaks up the spiral arms before they are 
virialized.

We do not insist that our simple model gives a fully correct description
of the mechanism through which $Q$ controls the amplitude of spiral 
arms, but it is clear from Figure \ref{fig:Q_tp} that $Q$ determines 
the amplitude. Thus, the spiral arms evolve in a 
self-regulated manner as follows.
Spiral arms grow from small density perturbations by the swing 
amplification \citep{Toomre81} to their maximum amplitudes
limited by $Q$. 
The spiral arms scatter disk stars, and as a result the velocity 
dispersion of the disk star increases. The heating rate, $dQ/dt$,
is proportional to the squared amplitudes of
spiral arms (see Equation (\ref{eq:dQdt2})).
This heating mechanism increases $Q$, and therefore 
the amplitude of spiral arms decreases. 
As a result, the heating rate decreases as $Q$ increases.
Through this evolution, the spiral arms become asymptotically 
faint as qualitatively shown by the black line in Figure \ref{fig:Q_tp}, 
but its timescale is comparable to the cosmological time, i.e., 
10 Gyr. 

\subsection{Effects of the initial $Q$ value and disk mass}
\begin{figure*}[htbp]
\epsscale{1.00}
\plotone{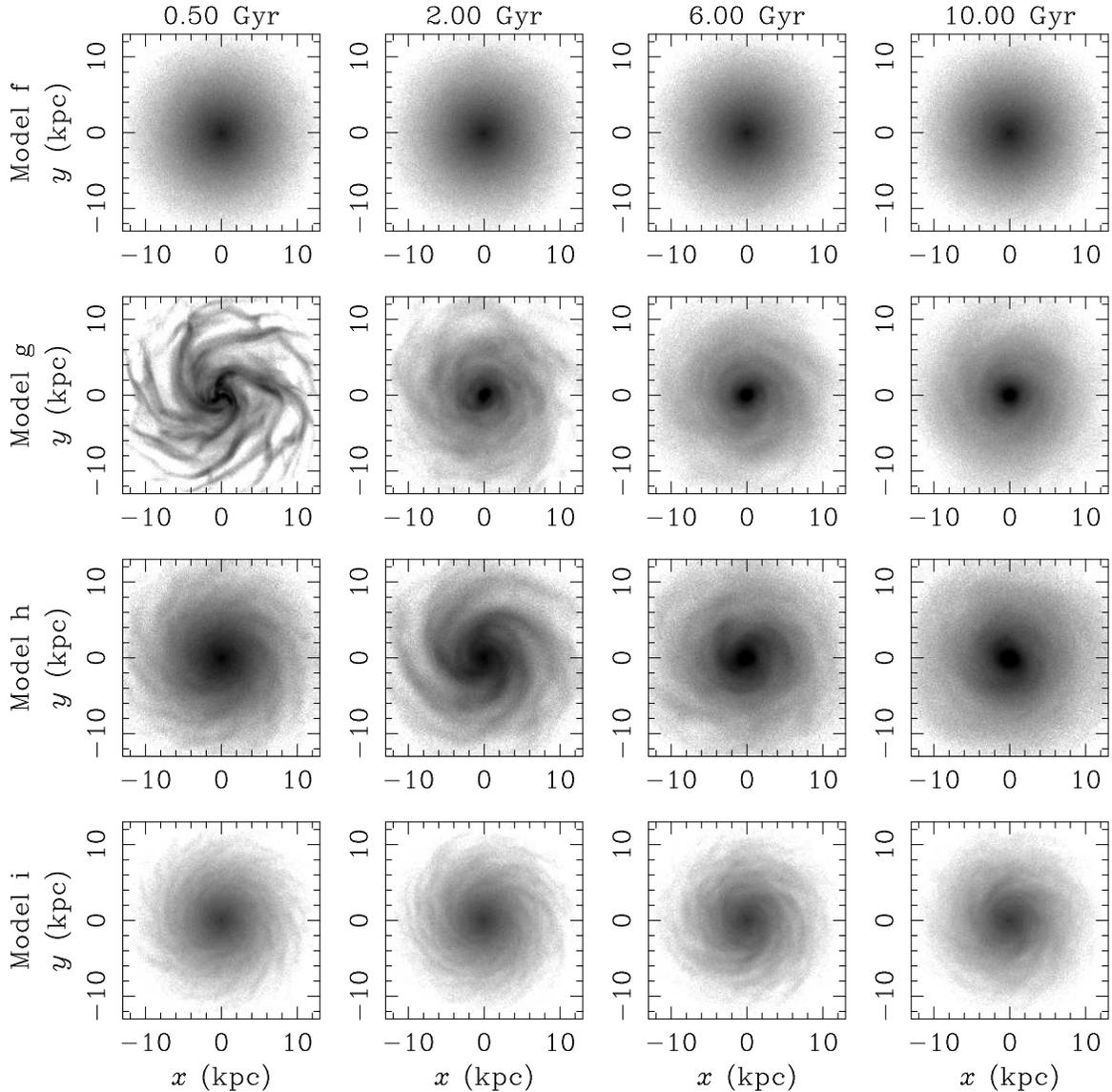}
    \caption{Snapshots of models f--i.}
 \label{fig:snapshot_all}
\end{figure*}

In order to see how the initial $Q$, and disk mass fraction 
affect the evolution and morphology of spiral arms,
we performed three additional runs: an unstable disk ($Q_0=0.5$; 
model g),
a massive disk ($M_{\rm d}/M_{\rm h} = 0.075$, model h), and a less massive 
disk ($M_{\rm d}/M_{\rm h} = 0.03$, model i).
In all models, the number of particles is $3\times 10^6$. 
Figure \ref{fig:snapshot_all} shows the snapshots of
models f--i. Model f is a model with a large initial $Q$, 
$Q_0=1.8$ and shown in Section 3.2. We show it again as an example 
of a model with a large $Q_0$.

\begin{figure}[htbp]
\epsscale{1.0}
\plotone{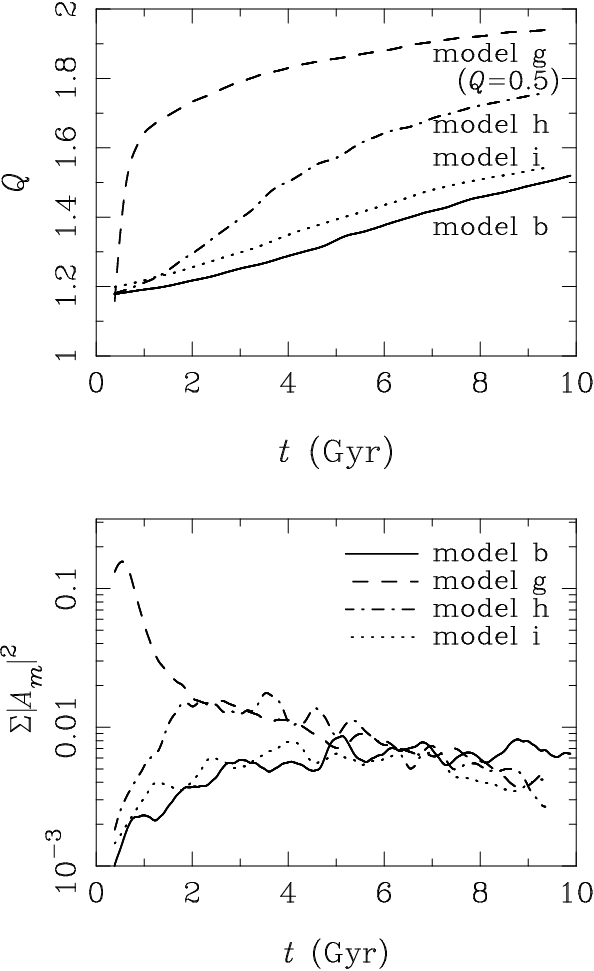}
   \caption{Time evolution of $Q$ averaged within 5--10 kpc (top) 
and total power, $\sum _{m=1}^{10}|A_m|^2$, averaged in 1 kpc at 7.5 kpc
(bottom) for models b, g, h, and  i, at $R=7.5$ (kpc).}
 \label{fig:3M_all}
\end{figure}

Model g is initially cold and unstable, therefore strong 
spiral arms develop in the first 0.5 Gyr. As shown in top panel of 
Figure \ref{fig:3M_all}, the disk is soon heated up to 
$Q \sim 1.6$, and the $Q$ keeps increasing.  
As expected from our theory in Section 3.3, the amplitude of spirals then 
decreases quickly, and they are very weak at 
$t=6$ Gyr. The final density and velocity profiles are quite different 
from original ones. 
In model f, whose disk is initially hot ($Q_0= 1.8$), 
spiral arms do not develop, and therefore
$Q$ (or the velocity dispersion) stays nearly constant 
(see the left panels of Figure \ref{fig:three}).

Models h and i have the same parameters as those of 
model b (standard model) except the disk mass ratio to halo, 
$M_{\rm d}/M_{\rm h}$ . \citet{CF85} showed that the number of spiral arms 
in numerical simulation is consistent with that predicted by 
the swing amplification theory \citep{Toomre81}, and massive disks 
have a smaller number of spiral arms. Our results are consistent with
this previous result. 
We can estimate the number of spiral arms as follows. 
The swing amplification is characterized by a parameter, 
$X\equiv {k_{\rm cr}R}/{m} = {\kappa^2R}/{2\pi G\Sigma m}$,
where $k_{\rm cr}$ is the critical wavenumber \citep{BT08}.
Spiral arms develop most effectively when $1<X<2$ 
\citep{Toomre81}. 
Therefore, we can estimate the dominating number of spiral arms, $m$, 
as
\begin{eqnarray}
m = \frac{\kappa ^2R}{2\pi G \Sigma X}
  \simeq \frac{\kappa ^2R}{4\pi G \Sigma},
\label{eq:m}
\end{eqnarray}
where we adopted $X\simeq2$.
In the case of model b, we obtain  $m=6$ at 8 kpc, 
which roughly agrees with the result of the simulation. 
This estimate is applicable for other models with different
disk mass fractions. 
Since the halo mass, $M_{\rm h}$, is fixed in our models, the surface density,
$\Sigma$, is proportional to the disk mass fraction, $M_{\rm d}/M_{\rm h}$. 
The numbers of spiral arms of models h and i are estimated as
$m=4$ and $m=9$ at 8 kpc, and they also agree with the results of the 
simulations.

\begin{figure*}[htbp]
\epsscale{1.00}
\plottwo{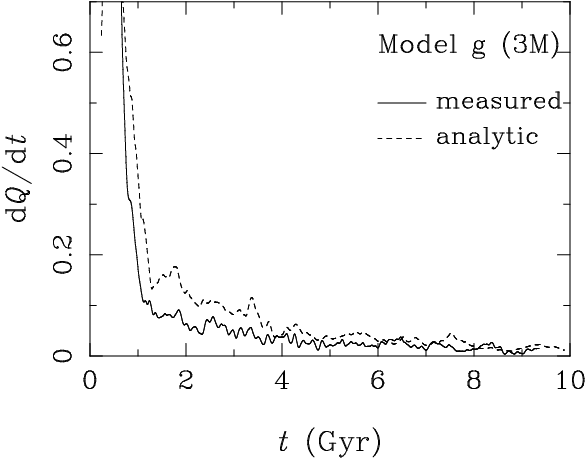}{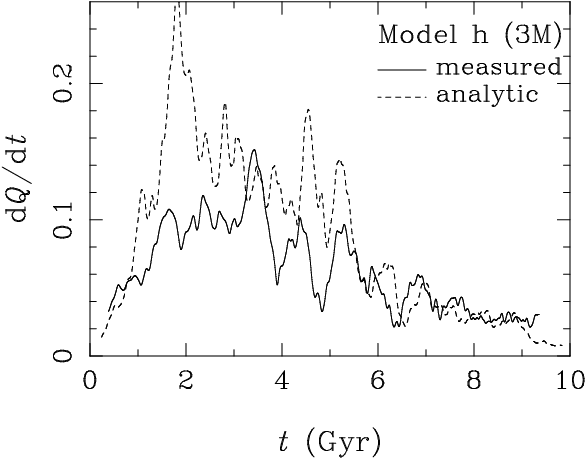}
 \caption{Comparison of $dQ/dt$ between measured and analytic results 
 for models g (left) and h (right), 
$N=3$M $R=7.5 \pm 0.5$ kpc.\label{fig:comp_gh_3M}}
\end{figure*}

Figure \ref{fig:3M_all} shows the time evolution of $Q$ averaged 
within 5--10 kpc (top) and the total power at $R=7.5$ kpc (bottom), 
respectively. As was the case with models a--f, 
the large total powers correspond to the rapid increase of $Q$.  
We investigated the time evolutions of $dQ/dt$ from the simulations and 
evaluated them using Equation (\ref{eq:dQdt}) in the same way as 
models a--f. Figure \ref{fig:comp_gh_3M} shows the results. 
The analytic results again agree well with the simulations in these models.

In fact, the evolution of these models are qualitatively similar, but 
different in details. The evolution of model h, which has a 
more massive disk, is faster than that of model b (our standard model). 
In model h, the total power of the spiral arms grows to 
$\sum |A_m|^2\sim 0.02$, at $t\sim 2$--4 Gyr, whereas it is 
$3\times 10^{-3}$ in model b. This difference causes the faster decay 
of the amplitude in model h (see bottom panel of Figure \ref{fig:3M_all}).
Furthermore, the number of spiral arms decreases
in model h. Initially, the number is around four (see 2.0 Gyr
in Figure \ref{fig:snapshot_all}), but three at 6.0 Gyr.
In model h, the effective angular-momentum transport occurs due to 
asymmetric 
structures in the disk and the surface density of the inner region 
increases. This reduces the number of the spiral arms.

\subsection{Effects of the number of particles}
As we showed in Figures \ref{fig:snapshot1_30M} -- \ref{fig:snapshot1_300k}, 
while the spiral arms survive for more 
than 10 Gyr in $N=3$M, 9M, and 30M models, this is not the case in 
$N=300$k and 1M models. Although the disks in $N=300$k 
models become featureless after $t=6$ Gyr (Figure \ref{fig:snapshot1_300k}), 
this is not mainly by the effect of the two-body relaxation. 
We can estimate the heating rate due to the two-body relaxation from the
result of model f, where spiral arms do not develop, the relative
heating rate for model f with $N=300$k at $Q=1.8$ is around 
0.5\%/Gyr. For a disk, the relative heating rate is proportional to 
$Q^{-4}$. Thus, for $Q=1.2$ and $N=300$k, the heating rate due to the 
two-body relaxation is $\sim 3$\%/Gyr, which is much smaller than the 
actual increase obtained from simulations. Heating due to
spiral arms is dominant even in $N=300$k models.

The major differences between $N=$3M--30M models and $N=$300k and
1M models are
(1) the time when the spiral arms reach their peak amplitudes and (2) the
peak amplitude themselves. The spiral arms initially grow from 
small perturbations of the density originated in the Poisson noise of 
disk stars and grow up to their maximum amplitude determined by $Q$. 
For the same initial $Q$, $N=300$k models reach
the maximum amplitudes much faster, and the values are large.  
A smaller number of particles generates a larger Poisson noise. Therefore,
spiral arms in $N=300$k and 1M models can reach its maximum amplitude faster 
than those in $N=3$M--30M models \citep{Sellwood10}. This means 
that the peak amplitude in $N=300$k models is larger than those in 
$N=3$M--30M models, because they can develop before the $Q$ 
becomes large. The larger peak amplitude
results in larger heating rate. Thus, the $Q$ of the $N=300$k 
models increases faster than that of the $N=3$M--30M models.

As shown in Figure \ref{fig:tp_b}, the evolution 
of the models with $N=$9M and 3M is not significantly different after 
$t\sim 2$ Gyr. In other words, once the amplitude of the spirals reaches 
a certain level ($\Sigma |A_m|^2 \sim 2\times10^{-3}$), 
the differences in the initial conditions are no longer important.
In section 3.1, we estimate the timescale of the initial exponential 
growth as around 0.4 Gyr. If we start from the noise level of real galaxies,
$\sim 10^{11}$ particles, it will take 1.6 Gyr longer than in the case of 
$N=9$M. This is still shorter than the cosmological timescale.

The two-body relaxation has serious effects on the heating of the disk 
only when 
the number of particle is quite small. For example, the number of 
particles used in the simulation of \citet{SC84} was only $2\times 10^4$. 
Since the heating timescale of our $N=300$k model is $\sim 10$ Gyr for 
$Q=1.0$, the relaxation time of their model would be $\sim 1$ Gyr. 
Thus, it  seems  that in \citet{SC84} spiral arms are 
weakened by the heating due to the two-body relaxation.

\section{Summary and Discussion}

\subsection{Summary}
We performed three-dimensional $N$-body simulations of pure stellar 
disks with spiral arms and investigated their dynamical evolution.
We confirmed that the spiral arms are transient and recurrent. 
Contrary to previous results, we found that spiral arms in pure stellar 
disks can survive for more than 10 Gyrs, when we use a sufficiently 
large number of particles. 
We also found that spiral arms of a stellar disk are self-regulated. 
The spiral arms grow by the swing amplification up to their maximum 
amplitudes determined by Toomre's $Q$ value at the moment.
The amplitude becomes smaller as $Q$ increases.
The spiral arms heat up the disk, or increase the velocity 
dispersion of stars, by scattering the disk stars. As a result,
$Q$ increases, and the amplitude of spirals is suppressed. 
We found that the heating rate, which is given by $dQ/dt$, 
is roughly proportional to the square of amplitudes. It means that 
the heating timescale becomes longer as $Q$ increases. 
Thus, the spiral arms heat up the stellar disk and increases $Q$, 
but at the same time the increase of $Q$ results in the decay of the 
spiral amplitudes and a smaller heating rate. This self-regulating 
relation among $Q$, spiral amplitudes, and the heating rate maintains 
the spiral arms for more than 10 Gyr.

In the case of the smaller number of particles ($N=300$k), 
however, the spiral arms become faint much faster than in the model 
with $N=3$M, 9M, and 30M. 
We found that the initial exponential growth of the
density perturbation depends on the number of particles (see 
Figure \ref{fig:tp_b}). Spiral arms initially grow from 
density perturbations originated from the Poisson noise of the initial 
condition through the swing amplification. A smaller number of 
particles results in a larger seed noise.
Therefore, spiral arms grow faster up to their maximum amplitude 
determined by $Q$. The rapid growth of the amplitude causes a rapid 
heating of the disk and also rapid decay of the spiral arms.

Our results show that the timescale of the initial exponential growth 
is $\sim 0.4$ Gyr. It means that even if we start with smooth 
disks $N\sim 10^{11}$ stars, the total power of arms reach the level of 
$\sim 2\times 10^{-3}$ in a few Gyr.
In real galaxies, moreover, the disks may be initially perturbed by
hierarchical mergers and/or GMCs.  
If the initial perturbation is comparable to or larger than this level, 
the spiral arms that look very similar to those in
real spiral galaxies are formed within a Gyr or less.
It is beyond the scope of the present paper to discuss
the amplitude and shape of the initial perturbation in a realistic 
situation, which should be ultimately studied in a cosmological context 
with a sufficiently high numerical resolution, e.g., the generated disk 
should be represented by larger number of particles to avoid numerical 
artifacts.

\subsection{Effects of Gas and Star Formation}

In this paper, we showed that the presence of gas is not essential 
in maintaining spiral arms. 
However, real spiral galaxies have gas and its effect is not negligible. 
Gas in galactic disks can work as both cooling and heating sources. 
The velocity dispersion of gas is smaller than that of stars because gas 
clumps lose their kinetic energy due to dissipation when they collide 
each other. 
As suggested by \citet{SC84}, new stars formed from the gas have smaller 
velocity dispersions than that of old disk stars because of the smaller
velocity dispersion of gas.
The new stars can dynamically cool the disk. Moreover, the existence of 
gas reduces the effective $Q$ of the disk \citep{JS84}.
On the other hand, the smaller $Q$ makes the disk dynamically 
unstable \citep{BR88} and it may cause faster heating of the disk. 
In addition, gas trapped into stellar spiral arms would cool due to the 
dissipation and its gravity may strengthen the amplitude of spiral arms. 
The larger amplitude of spiral arms will cause faster heating of the 
disk. However, even if the gas enhances the amplitude of spiral arms, the 
self-regulating mechanism that we suggested in this paper will determine 
the spiral amplitudes. Therefore, 
the disk would keep the spiral arms for a long time as in the case of 
pure stellar disks. In reality, the interaction between the gas and 
stellar spirals is complicated. We are now performing 
simulations with gas and will present the results elsewhere 
(Baba et al. in preparation).

\acknowledgments

The authors thank Hiroshi Daisaka for setting up GRAPE-7 and Shunsuke
Hozumi for useful discussions.
M.S.F. and T.R.S. are financially supported by Research Fellowships of 
JSPS for Young Scientist.
Calculations were done using the
GRAPE system at the Center for Computational Astrophysics of
the National Astronomical Observatory of Japan.

\appendix

\section{Effects of opening angle of treecode}

\begin{figure}[htbp]
\epsscale{0.5}
\plotone{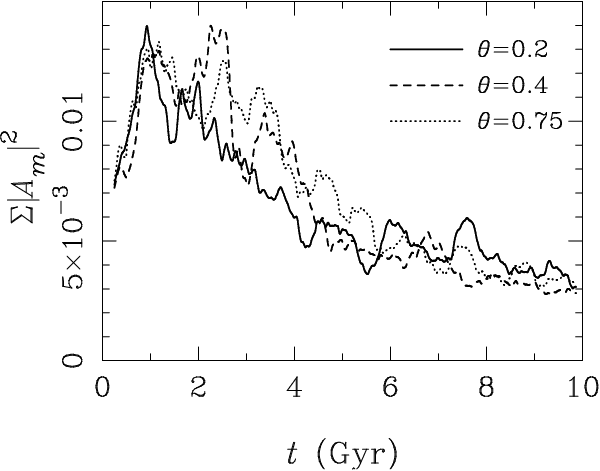}
 \caption{Evolutions of total powers at $R=7.5 \pm 0.5$ kpc for model b with 
$N=300$k. Solid, dashed, and dotted curves show the results with opening 
angles, $\theta = 0.2, 0.4$, and $0.75$, respectively. 
\label{fig:th_comp}}
\end{figure}

\begin{figure}[htbp]
\epsscale{0.5}
\plotone{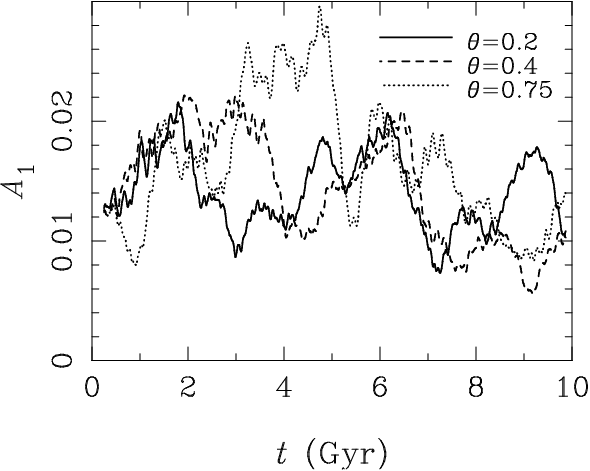}
 \caption{Time evolution of the amplitude of $m=1$, $A_1$, at 
$R=7.5 \pm 0.5$ kpc for model b with $N=300$k. 
Solid, dashed, and dotted curves show the results with opening 
angles, $\theta = 0.2, 0.4$, and $0.75$, respectively. 
\label{fig:th_m1_comp}}
\end{figure}

We used a treecode for calculation of forces. Since a treecode does not 
conserve the linear momentum of systems, the disks might drift during the 
simulations. Since we used a fixed background potential as a halo model,
the drift from the center of the halo might result in an artificial 
$m=1$ perturbation.
This effect depends on the opening angle of treecodes, $\theta$, which
is a parameter in the tree approximation. Smaller $\theta$ can achieve
more computational accuracy, but requires longer CPU time. 
If the drift is significant to induce the artificial perturbation, 
we expect larger amplitude of spirals developing in the later phase 
for larger $\theta$. In order to confirm this,
we performed simulations with $\theta = 0.2, 0.4$, and $0.75$ 
for model b with $N=300$k and investigated their convergence.
Figure \ref{fig:th_comp} shows the evolution of the total powers.
They do not show any clear differences. 
We also confirmed that the amplitudes of the $m=1$ mode do not grow 
throughout the simulations (see Figure \ref{fig:th_m1_comp}).
In the results in Section 3, we adopted $\theta = 0.4$ for all simulations.

\section{The time evolution of the Fourier amplitudes at $R<1$ kpc}
In this section, we show 
the time evolution of the Fourier amplitudes at $R<1$ kpc of the disk, 
which corresponds to the deviation of the density center of 
the disk from the center of the halo potential, in order to confirm that 
the drift of the density center of the disk is not significant.

\begin{figure}[htbp]
\epsscale{0.5}
\plotone{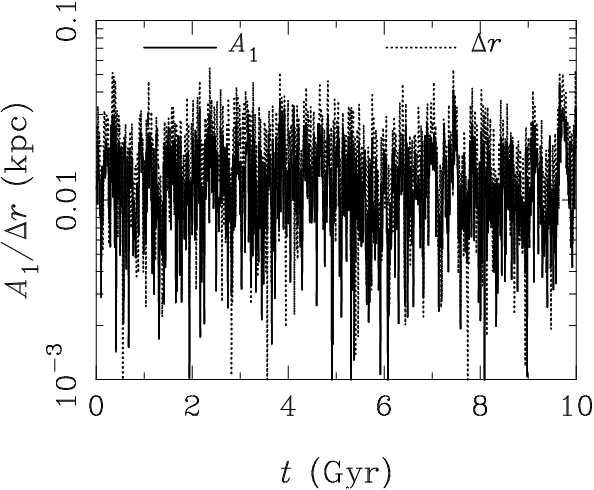}
 \caption{Time evolution of the amplitude of $m=1$, $A_1$, 
and the distance of the density center from the origin 
at $R<1$ kpc, $\Delta r$, for model b with $N=300$k and $\theta=0.4$. 
\label{fig:comp_A1_dc}}
\end{figure}

We investigated the deviation of the density center of the disk 
from the origin, which is the center of the halo potential, on the
$x$-$y$ plane. This directly shows the drift of the disk.
The dotted curve in Figure \ref{fig:comp_A1_dc} shows the distance of 
the density center from the origin at $R<1$ kpc for model b 
with $N=300$k, $\Delta r$.  The density center is calculated from the 
highest local density using the method of \citet{CH85}. Figure 
\ref{fig:comp_A1_dc} shows that
the deviation of the density center is as small as the softening 
length of 0.03 kpc and does not increase during the simulation.
The full curve in Figure \ref{fig:comp_A1_dc} shows the time evolution 
of the Fourier amplitudes of $m=1$, $A_1$. 
It shows that $A_1$ traces the deviation of the density center, $\Delta r$. 
The amplitude is due to the Poisson noise of the particle distribution. 
Hereafter, therefore, we adopt $A_1$ at 
$R<1$ kpc as the index of the deviation of the disk center from the 
halo center.

\begin{figure}[htbp]
\epsscale{1.0}
\plottwo{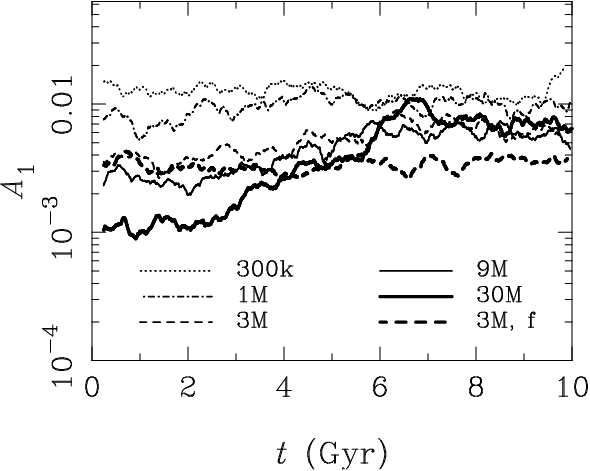}{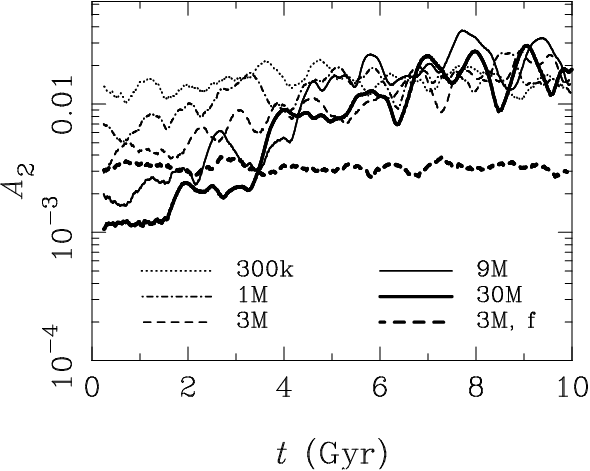}
\plottwo{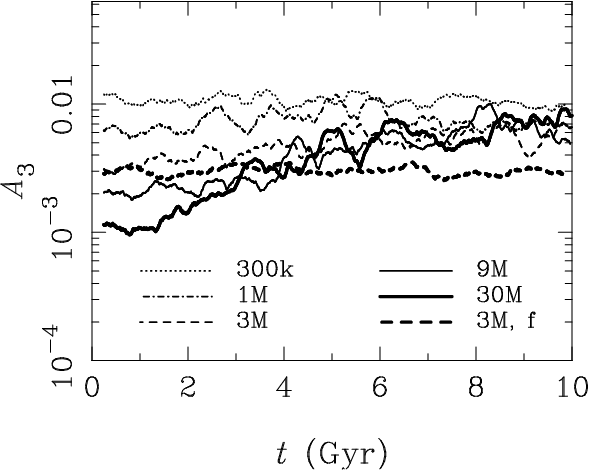}{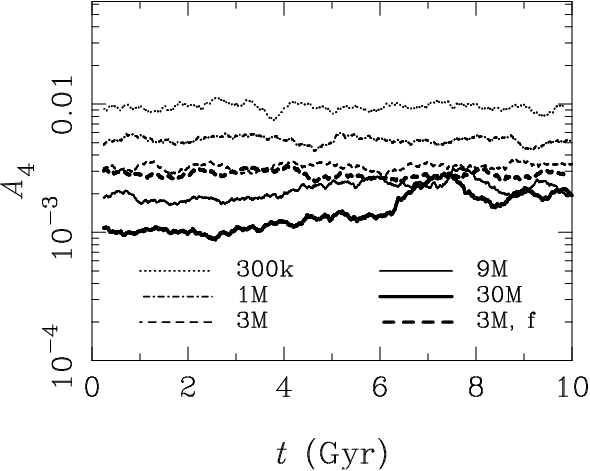}
\plottwo{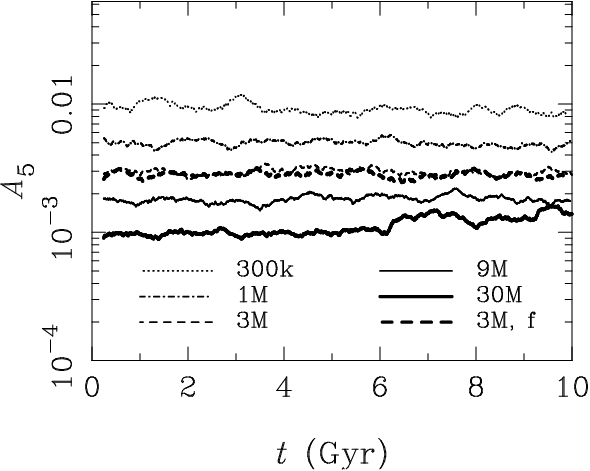}{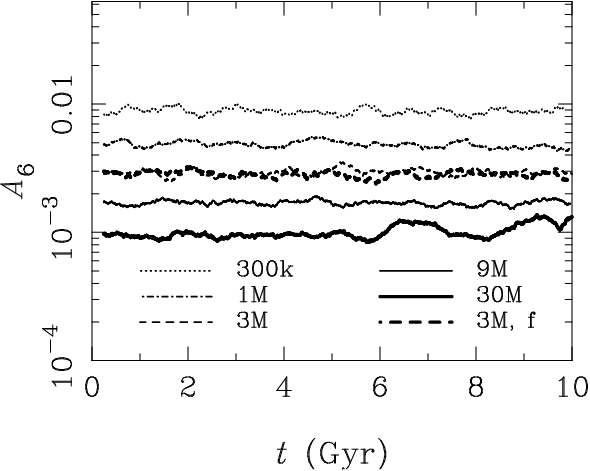}
 \caption{Time evolution of the Fourier amplitudes at $R<1$ kpc. 
\label{fig:amp_R1}}
\end{figure}

Figure \ref{fig:amp_R1} shows the Fourier amplitudes of $m=$1--6 at $R<1$
kpc for model b. These amplitudes are averaged over 0.5 Gyr. 
For comparison, we also plotted the result of model f with $N=3$M, in 
which spiral arms do not develop, because the initial $Q$ is large. 
Initially, all amplitudes show the fluctuations due to Poisson noise. In the 
case of small numbers of particles ($N<1$M), the amplitudes do not increase 
during the simulations. In the case of larger numbers of particles ($N>3$M), 
on the other hand, the amplitudes of $m=1$--3 modes increase after 
the spiral arms developed (after $\sim 2$ Gyr). We consider that this 
increase of the amplitudes at $R<1$ kpc is caused by the spiral arms  
developed in the outer part of the disk (e.g., at 7.5 kpc), because the 
amplitudes does not increase in the case of the disk without spiral 
arms (see model f in figure \ref{fig:amp_R1}). Moreover,
it is unlikely that these small increases of the amplitude at $R<1$ 
kpc affect the evolution of the spiral arms in the outer part of the disk. 
In model b with $N=30$M, for example, 
the amplitude grows to $\sim 30$ times as much as the initial fluctuation 
during the first 2 Gyr (see figure \ref{fig:tp_b}). After 2 Gyr, the 
amplitudes at $R<1$ kpc start to increase. 
If the increase of the amplitudes at $R<1$ kpc cause the evolution of 
the spiral arms of the outer part, the increase during 2--4 Gyr
at $R=7.5$ kpc must be larger than those of the first 2 Gyr. In fact,
however, the spiral amplitudes at $R=7.5$ kpc grow to only 2--3 times
during 2--4 Gyr. Therefore, we can safely conclude that the increase of the 
amplitudes at $R<1$ kpc is caused by the asymmetry of the self-developed 
spiral arms in the disk.

\newpage


\begin{thebibliography}{}
\bibitem[Athanassoula et al.(2005)]{Athanassoula05} Athanassoula, E., 
Lambert, J.~C., \& Dehnen, W.\ 2005, \mnras, 363, 496 

\bibitem[Baba et al.(2009)]{Baba09} Baba, J., Asaki, Y., 
Makino, J., Miyoshi, M., Saitoh, T.~R., \& Wada, K.\ 2009, \apj, 706, 
471 

\bibitem[Barnes \& Hut(1986)]{BH86} Barnes, J., \& Hut, P. 1986,
  \nat, 324, 446

\bibitem[Bertin \& Lin(1996)]{BL96} Bertin, G., \& Lin, C.~C.\ 1996, 
Spiral structure in galaxies a density wave theory, Publisher: 
Cambridge, MA MIT Press

\bibitem[Bertin 
\& Romeo(1988)]{BR88} Bertin, G., \& Romeo, A.~B.\ 1988, \aap, 195, 105 

\bibitem[Binney \& Tremaine(2008)]{BT08} Binney, J., \& Tremaine, S.\ 
2008, Galactic Dynamics: Second Edition, by James Binney and Scott 
Tremaine.~ISBN 978-0-691-13026-2 (HB).~Published by Princeton 
University Press, Princeton, NJ USA, 2008.

\bibitem[Bottema(2003)]{Bottema03} Bottema, R.\ 2003, \mnras, 
344, 358 

\bibitem[Carlberg 
\& Freedman(1985)]{CF85} Carlberg, R.~G., \& Freedman,
			     W.~L.\ 1985, \apj, 298, 486 
\bibitem[Carlberg 
\& Sellwood(1985)]{CS85} Carlberg, R.~G., \& Sellwood, 
J.~A.\ 1985, \apj, 292, 79 

\bibitem[{{Casertano} \& {Hut}(1985)}]{CH85}
{Casertano}, S., \& {Hut}, P. 1985, \apj, 298, 80

\bibitem[Donner 
\& Thomasson(1994)]{DT94} Donner, K.~J., \& Thomasson, M.\ 1994, \aap, 
290, 785 

\bibitem[Dubinski et al.(2009)]{DBS09} Dubinski, J., 
Berentzen, I., \& Shlosman, I.\ 2009, \apj, 697, 293 

\bibitem[Elmegreen 
\& Thomasson(1993)]{ET93} Elmegreen, B.~G., \& Thomasson, M.\ 1993, 
\aap, 272, 37 

\bibitem[Fuchs et al.(2005)]{Fuchs05} Fuchs, B., Dettbarn, C., \& Tsuchiya, T.\ 2005, \aap, 444, 1 


\bibitem[Goldreich \& Lynden-Bell(1965)]{GL65} Goldreich, P., \& 
Lynden-Bell, D.\ 1965, \mnras, 130, 97 

\bibitem[Hernquist(1993)]{H93} Hernquist, L.\ 1993, \apjs, 
86, 389 


\bibitem[Jog \& Solomon(1984)]{JS84} Jog, C.~J., \& Solomon, P.~M.\ 1984, 
\apj, 276, 114 

\bibitem[Julian \& Toomre(1966)]{JT66} Julian, W.~H., \& Toomre, A.\ 
1966, \apj, 146, 810 

\bibitem[Lin \& Shu(1964)]{LS64} Lin, C.~C., \& Shu, F.~H.\ 1964, \apj,
			     140, 646 

\bibitem[Kawai et al.(2006)]{Kawai06} Kawai, A., Fukushige, T., \& Makino, J.
2006, proceedings of SC06

\bibitem[Kokubo \& Ida(1992)]{KI92} Kokubo, E., \& Ida, S.\ 1992, 
\pasj, 44, 601 

\bibitem[Makino(1991)]{M91}
    Makino, J. 1991, \pasj, 43, 621

\bibitem[Makino (2004)]{M04}
    Makino, J. 2004, \pasj, 56, 521

\bibitem[Makino et al.(2007)]{Makino07} Makino, J., Hiraki, K., \& Inaba,
M. Proceedings of the 2007 ACM/IEEE, 2007


\bibitem[Mark(1976)]{Mark76} Mark, J.~W.-K.\ 1976, \apj, 206, 
418 


\bibitem[McMillan \& Dehnen(2007)]{MD07} McMillan, P.~J., \& Dehnen, 
W.\ 2007, \mnras, 378, 541 

\bibitem[Navarro et al.(1997)]{NFW97} Navarro, J.~F., Frenk, 
C.~S., \& White, S.~D.~M.\ 1997, \apj, 490, 493 


\bibitem[Sellwood(2000)]{Sellwood00} Sellwood, J.~A.\ 2000, \apss, 
272, 31 

\bibitem[Sellwood(2010)]{Sellwood10} Sellwood, J.~A.\ 2010, 
arXiv:1001.5430 

\bibitem[Sellwood 
\& Binney(2002)]{SB02} Sellwood, J.~A., \& Binney, J.~J.\ 2002, \mnras, 
336, 785 

\bibitem[Sellwood \& Carlberg(1984)]{SC84} Sellwood, J.~A., \& 
Carlberg, R.~G.\ 1984, \apj, 282, 61 

\bibitem[Sellwood 
\& Debattista(2009)]{SD09} Sellwood, J.~A., \& Debattista, V.~P.\ 2009,
			     \mnras, 398, 1279

\bibitem[Toomre(1969)]{Toomre69} Toomre, A.\ 1969, \apj, 158, 
899 

\bibitem[Toomre(1981)]{Toomre81} Toomre, A.\ 1981, Structure and 
Evolution of Normal Galaxies, 111

\bibitem[Toomre(1990)]{Toomre90} Toomre, A.\ 1990, Dynamics and 
Interactions of Galaxies, 292 

\bibitem[Toomre \& Kalnajs(1991)]{TK91} Toomre, A., \& Kalnajs, A.~J.\ 1991, Dynamics of Disc Galaxies, 341 


\end{thebibliography}
\end{document}